
\magnification=1200
\vskip 2cm
\hfill hep-th/9209040
\vskip 3cm

\centerline{\bf INTEGRABLE SYSTEMS AND CLASSIFICATION}
\centerline{\bf  OF 2-DIMENSIONAL
TOPOLOGICAL FIELD THEORIES.}
 \medskip

\centerline{B.Dubrovin\footnote{$^\dagger$}{On leave of absence from
Department Mechanics and Mathematics, Moscow State University,
119899 Moscow, Russia.}}
\medskip
\centerline{\sl SISSA}
\centerline{\sl via Beirut, 2}
\centerline{\sl I-34013 TRIESTE, Italy}
\vskip 2cm
{\bf Abstract.}
\medskip
In this paper we consider from the point of view of differential geometry
and of the theory of integrable systems the so-called WDVV equations as
defining relations of 2-dimensional topological field theory. A complete
classification of massive topological conformal field theories (TCFT)
is obtained in terms of monodromy data of an auxillary linear operator
with rational coefficients. Procedure of coupling of a TCFT to
topological gravity is described (at tree level) via certain integrable
bihamiltonian hierarchies of hydrodynamic type and their $\tau$-functions.
A possible role of bihamiltonian formalism in calculation of high genus
corrections is discussed. As a biproduct of this discussion new examples
of infinite dimensional Virasoro-type Lie algebras and their nonlinear
analogues are constructed. As an algebro-geometrical applications it is
shown that WDVV is just the universal system of integrable differential
equations (high order analogue of the Painlev\'e-VI) specifying periods of
Abelian differentials on Riemann surfaces as functions on moduli of
these surfaces.

\vskip 2cm

\centerline{Ref. S.I.S.S.A. 162/92/FM (September 92)}

\vfill\eject

This paper is an extended version of the talk given at the Workshop
\lq\lq Integrable Systems" (Luminy, July 1991). Also more recent results [39]
are
included. I dedicate it to the memory of J.-L.Verdier.
\medskip

{\bf Introduction.}
\medskip
A quantum field theory (QFT) on a $D$-dimensional manifold $M$ consists of:

1). a family of local fields $\phi_\alpha (x),~x\in M$ (functions or
sections of a fiber bundle over $M$). A metric $g_{ij}(x)$ on $M$
usualy is one of the fields (the gravity field).

2). A Lagrangian $L=L(\phi , \phi_x, ...)$. Classical field theory is
determined by the Euler -- Lagrange equations
$${\delta S\over \delta\phi_\alpha (x)} = 0,~~S[\phi ]=\int L(\phi ,
\phi_x, ...).$$

3). Procedure of quantization usualy is based on construction of an
appropriate  path integration
measure $[d\phi ]$. The partition function is a result of the path integration
over the space of all fields $\phi (x)$
$$Z_M=\int [d\phi ]e^{-S[\phi ]}.$$
Correlation functions (non normalized) are defined by a similar path
integral
$$<\phi_\alpha (x)\phi_\beta (y)\dots >_M = \int [d\phi ]\phi_\alpha
(x)\phi_\beta (y)\dots e^{-S[\phi ]}.$$
Since the path integration measure is almost never well-defined (and
also taking in account that different Lagrangians could give
equivalent QFT) an old idea of QFT is to construct a self-consistent
QFT by solving a system of differential equations for correlation
functions. These equations were scrutinized in 2D
conformal field theories where D=2 and Lagrangians are invariant with
respect to conformal transformations
$$\delta g_{ij}(x)=\epsilon g_{ij}(x), ~~\delta S = 0.$$
This theory is still far from being completed since complexity
(and, probably, nonintegrability) of the differential equations
determining correlators.

Here I will consider another class of solvable QFT: topological field
theories. These theories admit {\it topological invariance}: they are
invariant with respect to arbitrary change of the metric $g_{ij}(x)$
on $M$
$$\delta g_{ij}(x)={\rm arbitrary},~~\delta S=0.$$
On the quantum level that means that the partition function $Z_M$
depends only on topology of $M$. All the correlation functions also
are topological creatures: they depend only on the labels of operators
and on topology of $M$ but not on positions of operators
$$<\phi_\alpha (x)\phi_\beta (y)\dots >_M \equiv
<\phi_\alpha\phi_\beta\cdots >_M.$$
The simplest example is 2D gravity with the Hilbert -- Einstein action
$$S = \int R\sqrt{g} d^2x={\rm Euler~characteristic~of~}M.$$
There are two ways of quantization of this functional. The first one
is based on an appropriate discrete version of the model ($M \to $
polihedron). This way leads to considering matrix integrals of the
form [55]
$$Z_N(t) = \int_{X^*=X}\exp\{-{\rm tr}(X^2+t_1X^4+t_2X^6+\dots )dX$$
where the integral should be taken over the space of all $N\times N$
Hermitian matrices $X$. Here $t_1$, $t_2$ ... are called coupling
constants. A solution of 2D gravity [1] is based on the observation that
after an appropriate limiting procedure $N\to\infty$
(and a renormalization of $t$) the limiting
partition function coincides with $\tau$-function of the
KdV-hierarchy.

Another approach to 2D gravity is based on an appropriate
supersymmetric extension of the Hilbert -- Einstein Lagrangian [2]. This
reduces the path integral over the space of all metrics $g_{ij}(x)$ on
a surface $M$ of the given genus $g$ to an integral over the
finite-dimensional space of conformal classes of these metrics, i.e.
over the moduli space ${\cal M}_g$ of Riemann surfaces of genus $g$.
Correlation functions of the model are expressed via intersection
numbers of some cycles on the moduli space [2-4, 46, 49]
$$\phi_\alpha \leftrightarrow c_\alpha \in H_*({\cal M}_g),~~\alpha\in {\bf
N}$$
$$<\phi_\alpha\phi_\beta\dots >_g= \# (c_\alpha\cap c_\beta\cap\dots
)$$
(here the supscript $g$ means correlators on a surface of genus $g$).
This approach is often called {\it cohomological field theory}.

It was conjectured by Witten that the both approaches to 2D quantum
gravity should give the same results. This conjecture was proved by
Kontsevich [42-43]. He showed that the generating function
$$F(t) = \sum_g\sum_{\alpha ,\beta \dots}{t_1^\alpha \over \alpha ! }
{t_2^\beta \over \beta !}\dots <\phi_\alpha\phi_\beta\dots >_g$$
(the free energy of 2D gravity) is logarythm of $\tau$-function of a
solution of the KdV hierarchy (this was the original form of the
Witten's conjecture). The $\tau$-function is specified by the string
equation (see eq. (3.16b) below).

Other examples of 2D TFT
constructed in [2-6, 8-9, 46-49, 57] proved that these could have important
mathematical applications, probably being the best tool for treating
sophisticated topological objects.
In these examples correlators can be expressed via intersection numbers
on moduli spaces (or their coverings [48]) of holomorphyc maps of
Riemann surfaces to a complex (or even almost complex) variety ({\it
topological
sigma-models} [2]) or via intersection form of a singularity in the catastrophe
theory ({\it topological Landau -- Ginsburg models} [9, 7]; see also [10]).
(We do not discuss here interesting relations between these models.)
This gives rise to the following
\medskip
{\bf Problems}. What could be an intrinsic origin of integrability in 2D TFT?
How one can classify 2D TFT? Is it possible to find an analogue of the KdV
hierarchy for calculating  the partition function of a given TFT model?
\medskip
In this paper an approach to these problems is proposed being based on
differential geometry and
on the theory of classical integrable systems of KdV type. Main
ingredient of my approach is  Hamiltonian formalism of
integrable hierarchies of KdV type (see, e.g., [21, 25, 29]) and, especially,
Hamiltonian analysis of semi-classical limits of these systems [18-21].

Let me start with considering {\it matter sector} of a 2D topological
field theory. That means that the set of local fields $\phi_1(x),\dots
, \phi_n(x)$
(the so-called {\it primary fields} of the model)
does not contain the metric. (Afterwards one should
integrate over the space of metrics. This should give rise to a
procedure of {\it coupling to topological gravity} that will be
described below.) Then the correlators of the fields $\phi_1(x)$, ...,
$\phi_n(x)$ obey very simple algebraic axioms (a consequence [51] of
the general Atiyah's axioms [50] of a topological field theory).

Let
$$\eta_{\alpha\beta}=<\phi_\alpha\phi_\beta >_0$$
(0 means genus zero correlator),
$$c_{\alpha\beta\gamma}=<\phi_\alpha\phi_\beta\phi_\gamma >_0$$
Then

1) These tensors are symmetric and $\det (\eta_{\alpha\beta})\neq 0$. I
will use the tensor $\eta_{\alpha\beta }$ and the inverse
$(\eta^{\alpha\beta})=
(\eta_{\alpha\beta})^{-1}$
for lowering and raising indices.

2) $c_{\alpha\beta}^\gamma =
\eta^{\gamma\epsilon}c_{\alpha\beta\epsilon}$ is a tensor of structure
constants of a commutative associative algebra $A$ with a unity. That
means that for a basis $e_1$, ..., $e_n$ in $A$ the multiplication law
has the form
$$e_\alpha e_\beta =c_{\alpha\beta}^\gamma e_\gamma .$$
(We will normalise a basis in such a way that $e_1$ = the unity of
$A$. So $c_{1\alpha}^\beta =\delta_\alpha^\beta$.)

3) Let $H = \eta^{\alpha\beta}e_\alpha e_\beta \in A$. Then for
correlators of genus $g$ the following formula holds
$$<\phi_\alpha\dots \phi_\gamma >_g = <e_\alpha\dots e_\gamma ,
H^g>.$$

On this way
\medskip

Topologicaly invariant Lagrangian $\to$ correlators of local physical
fields
\medskip
\noindent we lose too much relevant information.
 To
capture more information on a topological Lagrangian we will consider
a topological field theory together with its deformations preserving
topological invariance
$$L\to L + \sum t^\alpha L_\alpha^{(pert)}$$
($t^\alpha$ are coupling constants). Here we use ideas and results of
[8, 51]. In these papers it was proposed a
general construction of a class of 2D TFT
by twisting of N=2 superconformal field theories. So-called {\it topological
conformal field theories} (TCFT) are obtained by this procedure. For any TCFT
with $n$ local observables (primary operators) it was constructed a
{\it canonical}
$n$-parameter deformation preserving topological invariance. All the
correlators of the primary fields $\phi_1$, ..., $\phi_n$ in
the perturbed TCFT now depend on coupling parameters
$t_1$, ..., $t_n$. This dependence is not arbitrary but obeys the
following eguations:

1). $\eta_{\alpha\beta} \equiv {\rm const~in}~t$

2). $c_{1\beta}^\alpha \equiv \delta_\beta^\alpha$

3). $c_{\alpha\beta\gamma} = {\partial^3F(t)\over \partial t^\alpha
\partial t^\beta \partial t^\gamma}$ for some function $F(t)$ (primary
free energy).
\medskip
Equations of associativity give a system of nonlinear PDE for $F(t)$
$${\partial^3F(t)\over \partial t^\alpha
\partial t^\beta \partial t^\lambda}
\eta^{\lambda\mu}
{\partial^3F(t)\over \partial t^\mu
\partial t^\gamma \partial t^\sigma}
=
{\partial^3F(t)\over \partial t^\alpha
\partial t^\gamma \partial t^\lambda}
\eta^{\lambda\mu}
{\partial^3F(t)\over \partial t^\mu
\partial t^\beta \partial t^\sigma}\eqno(0.1)$$
with the constraint
$$
{\partial^3F(t)\over \partial t^1
\partial t^\alpha \partial t^\beta} = \eta_{\alpha\beta}.\eqno(0.2)$$
These equations were called in [39] {\it Witten -- Dijkgraaf --
E.Verlinde -- H.Verlinde} (WDVV) equations.
In fact in TCFT one should assume invariance of a solution with respect to
scaling transformations of the form
$$t^\alpha \mapsto c^{1-q_\alpha}t^\alpha$$
$$\eta_{\alpha\beta}\mapsto c^{q_\alpha +q_\beta -
d}\eta_{\alpha\beta}$$
$$c_{\alpha\beta} \mapsto c^{q_\alpha +q_\beta -
q_\gamma}c_{\alpha\beta}^\gamma$$
for some numbers $q_\alpha$ ({\it charges} of the fields
$\phi_\alpha$) and $d$ ({\it dimension} of the model).

My program now is:

1. To classify 2D TFT as solutions of WDVV
equations, and

2. For any solution of WDVV (I recall that this
describes the matter sector of a TFT model) to construct
(i.e., to calculate partition function and correlators) a complete TFT
model (coupling of the given matter sector to topological gravity).

The problem 1 was investigated in [39]. A Lax pair for the WDVV
equations was constructed. For the so-called {\it massive} TCFT models
where the Fr\"obenius algebra for almost all $t$ has no nilpotents it
was shown that solutions of the WDVV equations form a
${n(n-1)\over 2}+1$-dimensional family (where $n$ is the
number of primaries). It
turns out that the WDVV equations for massive case are equivalent to
equations of isomonodromy deformations of an ordinary linear
differential operator with rational coefficients. These isomonodromy
deformation equations coincide with the Painlev\'e-VI equation (for
$n=3$) and with high order analogues of the Painlev\'e-VI for $n>3$. Monodromy
data (i.e. Stokes matrices) of the operator with rational
coefficients serve as parameters of massive TCFT-models.

Concerning the second problem my conjecture is that the set of
solutions of WDVV parametrizes a big class of hierarchies of
1+1-integrable systems. All the wellknown hierarchies are in this
class but they are only isolated points in it.

The basic idea of construction of these integrable hierarchies comes
from the standard in quantum field theory Feynmann diagram expansion
machinery. In 2D QFT it becomes a representation of partition function
and correlators as a sum of contributions from surfaces of different
genera $g$ ({\it genus expansion}). The idea is that the genus
expansion of the partition function should coincide with the small
dispersion expansion (see below) of the $\tau$-function of some integrable
hierarchy.

A first step on this way has been done in [39]: for any solution of
WDVV a hierarchy of integrable Hamiltonian
equations of hydrodynamic type was constructed
such that  $\tau$-function of a particular solution of this
hierarchy coincides with the genus zero approximation of the
correspondent TFT model coupled to gravity (see Section 3 below).
{}From the point of view of WDVV equations the hierarchy determines a family
of symmetries of the equation (0.1) (see below, Proposition 3.1). To
go further one should solve a non-standard \lq\lq inverse problem": to
reconstruct integrable hierarchy from the zero-dispersion limit of it.
Some examples of such a reconstruction are discussed in Sections 3, 4.
Probably, bihamiltonian formalism  could be useful to complete
solution of this problem.

Examples of solutions of WDVV and of corresponding integrable
hierarchies are described in Section 4. Almost all the known examples
are obtained as a result of analysis of the semiclassical
(particularly, dispersionless) limiting procedure in integrable
hierarchies of KdV type. More precisely, let
$$\partial_{t_k}y^a = f_k^a(y,\partial_xy,\partial_x^2y,\dots ),~
a=1,\dots ,l,~k=0, 1,\dots $$
be a commutative hierarchy of Hamiltonian integrable systems of the KdV type.
\lq\lq Hierarchy" means that the systems are ordered, say, by action of a
recursion operator. Number of recursions determine a level of a system in the
hierarchy. Systems of the level zero form a primary part of the hierarchy
(these correspond to the primary operators in TFT); others can be obtained
from the primaries by recursions. The hierarchy
posesses a rich family of finite-dimensional invariant manifolds. Some of
them can be found in a straightforward way; one needs to apply sophisticated
algebraic geometry methods [28] to construct more wide class of invariant
manifolds. Any of these manifolds after an extension to complex domain
turns out to be fibered over some base $M$
(a complex manifold of some dimension $n$) with $m$-dimensional tori as the
fibers (common invariant tori of the hierarchy). For $m=0$ $M$ is nothing
but the family of common stationary points of the hierarchy. For $m>0$ $M$ is
a moduli space of Riemann  surfaces of some genus $g$ with certain additional
structures: marked points, marked meromorphic function etc. These are the
families of finite-gap (\lq\lq $g$-gap") solutions of the hierarchy. The main
observation is that any such $M$ determines a solution of WDVV equation ($M$
is a Fr\"obenius manifold in the terminology of Section 1 below,
or the \lq\lq small phase space" of a TFT theory in the terminology of [3-4,
46]).
For $m=0$
and the set $M$ of stationary points of the Gelfand -- Dickey hierarchy
this essentialy follows from [8, 11]; for general case (including arbitrary
genera $g$) a construction of solution of WDVV was given in [13-14] (see also
recent preprint [44]).

To give an idea how an integrable Hamiltonian hierarchy of the above form
induces tensors $c_{\alpha\beta}^\gamma$, $\eta_{\alpha\beta}$ on a finite
dimensional invariant manifold $M$ I need to introduce the notion of
semiclassical limit of a hierarchy near a family $M$ of invariant tori
(sometimes it is called also a {\it dispersionless limit } or {\it Whitham
averaging} of the hierarchy; see details in [15-21]). In the simplest case of
the family of stationary solutions the semiclassical limit is defined as
follows: one should substitute in the equations of the hierarchy
$$x\mapsto\epsilon x = X,~ t_k\mapsto\epsilon t_k = T_k$$
and tend $\epsilon$ to zero. For more general $M$ (family of invariant tori)
one should add averaging over the tori. As a result one obtains a new
integrable Hamiltonian hierarchy where the dependent variables are coordinates
$v^1$, ..., $v^n$ on $M$ and the independent variables are the slow
variables $X$ and $T_0$, $T_1$, ... . This new hierarchy always has a form of a
quasilinear system of PDE of the first order
$$\partial_{T_k}v^p = c_k{_q^p}(v)\partial_Xv^q,~k=0, 1, \dots$$
for some matrices of coefficients $c_k{_q^p}(v)$. One can keep in mind the
simplest example of a semiclassical limit (just the dispersionless limit) of
the KdV hierarchy. Here $M$ is the one-dimensional family of constant
solutions of the KdV hierarchy. For example, rescaling the KdV one obtains
$$u_T=uu_X+\epsilon^2u_{XXX}$$
(KdV with small dispersion). After $\epsilon\to 0$ one obtains
$$u_T=uu_X.$$
The semiclassical limit of all the KdV hierarchy has the form
$$\partial_{T_k}u = {u^k\over k!}\partial_Xu,~k=0, 1,\dots .$$

A semiclassical limit of spatialy discretized hierarchies (like Toda
system) is obtained by a similar way. It still is a system of quasilinear
PDE of the first order.

Let us come back to determination of tensors $\eta_{\alpha\beta}$,
$c_{\alpha\beta}^\gamma$ on $M$. To introduce $\eta_{\alpha\beta}$
we need to use the Hamiltonian structure of the original hierarchy. A
semiclassical limit (or \lq\lq averaging") of this Hamiltonian structure
in the sense of general construction of S.P.Novikov and the author induces
a Hamiltonian structure of the semiclassical hierarchy: a Poisson bracket of
the form
$$\{ v^p(X),v^q(Y)\}_{{\rm semiclassical}} =
g^{ps}(v(X))[\delta_s^q\partial_X\delta (X-Y) -
\Gamma_{sr}^q(v)v_X^r\delta (X-Y)]$$
where $g^{pq}(v)$ are contravariant components of a metric on $M$ and
$\Gamma_{pr}^q(v)$ are the Christoffel symbols of the Levi-Civit\'a connection
for $g^{pq}(v)$ (the so-called {\it Poisson brackets of hydrodynamic type}).
(Strictly speaking the metric and the connection are defined on a real part of
$M$ that parametrizes smooth solutions of the original hierarchy with some
reality constraints. But the formulae for the metric and the connection admit
an extension onto all $M$.) From the general theory of Poisson brackets of
hydrodynamic type [18-21] one concludes
that the metric $g^{pq}(v)$ on $M$ should
have zero curvature. So local flat coordinates $t^1$, ..., $t^n$ on $M$ exist
such that the metric in this coordinates is constant
$${\partial t^\alpha\over\partial v^p}{\partial t^\beta\over\partial v^q}
g^{pq}(v) = \eta_{\alpha\beta} = {\rm const}.$$
The Poisson bracket $\{ ~,~\}_{{\rm semiclassical}}$ in these coordinates has
the form
$$\{ t^\alpha (X),t^\beta (Y)\}_{{\rm semiclassical}} = \eta^{\alpha\beta}
\delta '(X-Y).$$
The tensor $(\eta_{\alpha\beta})=(\eta^{\alpha\beta})^{-1}$ together with the
flat coordinates $t^\alpha$ is the first part of a structure we want to
construct. (The flat coordinates $t^1$, ..., $t^n$ can be expressed
via Casimirs of the original Poisson bracket and action variables and wave
numbers along the invariant tori - see details in [18-21].)

To define a tensor $c_{\alpha\beta}^\gamma (t)$ on $M$ (or, equivalently,
the \lq\lq primary free energy" $F(t)$) we need to use a semiclassical limit
of the $\tau$-function of the original hierarchy [11, 53-54, 61]
$$\log\tau_{{\rm semiclassical}}(T_0,T_1,\dots ) =
\lim_{\epsilon\to 0} \epsilon^{-2}\log\tau (\epsilon t_0,\epsilon t_1,
\dots ).$$
Then
$$F = \log \tau_{{\rm semiclassical}}$$
where $\tau_{{\rm semiclassical}}$ should be considered as a function only
on $n$ of the slow variables of the same level.
(To satisfy the normalization (0.2) one should choose properly these $n$
slow variables and normalize values of others. This specifies uniquely the
semiclassical $\tau$-function.)
The semiclassical $\tau$-function as the function of all slow variables
coincides with the tree-level partition function of the matter sector
$\eta_{\alpha\beta}$, $c_{\alpha\beta}^\gamma$ coupled to topological gravity.

Summarizing, we can say that a structure of Fr\"obenius manifold
(i.e., a solution of WDVV) on an invariant manifold $M$ of an integrable
Hamiltonian hierarchy is induced by a semiclassical limit of the Poisson
bracket of the hierarchy and of the $\tau$-function of the hierarchy. So the
above conjecture can be reformulated as follows: WDVV equations just specify
the semiclassical limits of $\tau$-functions of Hamiltonian integrable
hierarchies.

I do not consider in this paper one more type of integrable systems
being involved in 2D TFT: the so-called equations of {\it
topological-antitopological fusion} proposed in [40]. These equations
describe the ground state metric on a given 2D TFT model. See [45]
about the theory of integrability of these equations. An interesting
relation of these equations to the theory of harmonic maps also was
found in [45].
\medskip
{\bf 1. Geometry of Fr\"obenius manifolds.}
\medskip
I recall that $A$ is called a Fr\"obenius algebra (over $\bf R$
or $\bf C$) if it is a commutative associative algebra with a unity and
with a nondegenerate invariant inner product
$$<ab,c>=<a,bc>.\eqno(1.1)$$
If $e$ is the unity of $A$ then the invariant inner product on $A$ can be
written in the form
$$<a,b>=\omega_e(ab)\eqno(1.2a)$$
where
$$\omega_e(a)=<e,a>.\eqno(1.2b)$$
Moreover, for any linear functional $\omega\in A^*$ the inner product
$$<a,b>_\omega =\omega (ab)\eqno(1.3)$$
is invariant. It is nondegenerate for generic $\omega$. Any invariant
inner product on a finite-dimensional Fr\"obenius algebra $A$ (only
finite-dimensional algebras will be considered) can be represented in the form
(1.3).

If $e_i,~ i=1,\dots ,n$ is a basis in $A$ then the structure of Fr\"obenius
algebra is specified by the coefficients $\eta_{ij}$, $c_{ij}^k$ where
$$<e_i,e_j>=\eta_{ij}\eqno(1.4a)$$
$$e_ie_j=c_{ij}^ke_k\eqno(1.4b)$$
(summation over repeated indices will be assumed). The matrix $\eta_{ij}$ and
the structure constants $c_{ij}^k$ satisfy the following conditions:
$$\eta_{ji}=\eta_{ij},~~\det (\eta_{ij})\neq 0\eqno(1.5a)$$
$$c_{ij}^sc_{sk}^l=c_{is}^lc_{jk}^s\eqno(1.5b)$$
(associativity),
$$c_{ijk}=\eta_{is}c_{jk}^s=c_{jik}=c_{ikj}\eqno(1.5c)$$
(commutativity and invariance of the inner product). If $e=(e^i)$ is the unity
of $A$ then
$$e^sc_{sj}^i=\delta_j^i\eqno(1.5d)$$
(the Kronecker delta).

1-dimensional Fr\"obenius algebras are parametrized by 1 number (length
of the unity). Any semisimple $n$-dimensional Fr\"obenius algebra is
isomorphic to the direct sum of $n$ one-dimensional Fr\"obenius algebras
$$f_if_j=\delta_{ij}f_i,~~<f_i,f_j>=\eta_{ii}\delta_{ij}.\eqno(1.6)$$
Moreover, any Fr\"obenius algebra without nilpotents is a semisimple one.

Let us consider a particular class of deformations of Fr\"obenius
algebras.

{\bf Definition 1.1.} A  manifold $M$ is called
{\it quasi-Fr\"obenius} if it is equipped with three tensors
$c=(c_{ij}^k(x))$, $\eta =(\eta_{ij}(x))$, $e=(e^i(x))$ satisfying (1.5)
 for any $x\in M$.

In other words these three tensors provide a structure of Fr\"obenius
algebra in the space of smooth vector fields $Vect(M)$ over the ring
${\cal F}(M)$ of smooth functions on $M$:
$$[v\cdot w]^k(x)=c_{ij}^k(x)v^i(x)w^j(x),\eqno(1.7a)$$
$$<v,w>(x)=\eta_{ij}(x)v^i(x)w^j(x)\eqno(1.7b)$$
for any $v,~w\in Vect(M)$.

Complex quasi-Fr\"obenius manifolds are defined in a similar way
but the tensors $c$, $\eta$, $e$ should be holomorphic. They
provide a structure of Fr\"obenius algebra in the space of
holomorphic vector fields over the ring of holomorphic functions.

Informaly speaking, $n$-dimensional
quasi-Fr\"obenius manifolds are $n$-parameter
deformations of $n$-dimensional Fr\"obenius algebras. For any $x\in M$
the tangent space $T_xM$ is a Fr\"obenius algebra with the structure
constants $c_{ij}^k(x)$, invariant inner product $\eta_{ij}(x)$, and
unity $e^i(x)$.

As it was explained above, in physical applications there are
additional restrictions for quasi-Fr\"obenius manifolds.

{\bf Definition 1.2.} A quasi-Fr\"obenius $M$ is called {\it Fr\"obenius
manifold} if the invariant metric
$$ds^2=\eta_{ij}(x)dx^idx^j\eqno(1.8a)$$
is flat, the unity vector field $e$ is covariantly constant
$$\nabla e=0\eqno(1.8b)$$
(here $\nabla$ is the Levi-Civit\'a connection for $ds^2$) and
the tensor
$$\nabla_z<u\cdot v,w>\eqno(1.8c)$$
is symmetric in the vectors $u$, ..., $z$.

Localy Fr\"obenius manifolds are in 1-1
correspondence with solutions of WDVV equations (i.e., with 2D TFTs).
Indeed, for the flat metric (1.8a) localy flat coordinates $t^\alpha$ exist
such that the metric is constant in these coordinates, $ds^2 =
\eta_{\alpha\beta}dt^\alpha dt^\beta$, $\eta_{\alpha\beta} = {\rm const}$.
The covariantly constant vector field $e$ in the flat coordinates has
constant components; using a linear change of the coordinates one can
obtain $e^\alpha = \delta_1^\alpha$. The tensor $c_{\alpha\beta\gamma}(t)$
in these coordinates satisfies the condition
$$\partial_\delta c_{\alpha\beta\gamma} =
\partial_\gamma c_{\alpha\beta\delta}.\eqno(1.9a)$$
This means that $c_{\alpha\beta\gamma}(t)$ can be represented in the form
$$c_{\alpha\beta\gamma}(t) = \partial_\alpha\partial_\beta\partial_\gamma
F(t)\eqno(1.9b)$$
for some function $F(t)$ satisfying the WDVV equations.

The first step in solving WDVV is to obtain a \lq\lq Lax pair" for
these equations. The most convenient way is to represent them as the
compatibility conditions of an overdetermined linear system depending on a
spectral parameter $\lambda$.

{\bf Proposition 1.1.} {\it A quasi-Fr\"obenius manifold is Fr\"obenius iff
the unity
$e$ is covariantly constant and the pencil of connections
$$\tilde\nabla_u(\lambda )v=\nabla_uv+\lambda u\cdot v,~~u,~v\in Vect(M)
\eqno(1.10)$$
is flat identicaly in $\lambda$.}

Flatness of the pencil of connections (1.10) is equivalent to flatness of the
metric $\eta$ and to the equation
$$\nabla_u(v\cdot w)-\nabla_v(u\cdot w)
+u\cdot\nabla_vw-v\cdot\nabla_uw=[u,v]\cdot w\eqno(1.11)$$
for any three vector fields $u$, $v$, $w$. Here $[u,v]$ means the commutator
of the vector fields. This equation is equivalent to the symmetry of the
tensor (1.8c).

{\bf Corollary.} {\it WDVV is an integrable system.}

Indeed, WDVV is equivalent to compatibility of the following linear system
$$\tilde\nabla_\alpha (\lambda )\xi =0,~~\alpha =1,...,n,\eqno(1.12a)$$
(here $\xi$ is a covector field), or, equivalently, in the flat coordinates
$t^\alpha$
$$\partial_\alpha\xi_\beta = \lambda c_{\alpha\beta}^\gamma (t)\xi_\gamma.
\eqno(1.12b)$$
Compatibility of the system (1.12) (identicaly in the spectral parameter
$\lambda$) together with the symmetry of the tensor
$c_{\alpha\beta\gamma}=\eta_{\alpha\epsilon}c_{\beta\gamma}^\epsilon$
is equivalent to WDVV.

It turns out that symmetries of Fr\"obenius manifolds play an important
role in geometrical foundation of TFT. We start with the notion of
 {\it algebraic
symmetry} of a Fr\"obenius manifold.

{\bf Definition 1.3.} A diffeomorphism $f:M\rightarrow M$ of a Fr\"obenius
manifold is called algebraic symmetry if it preserves the multiplication
law of vector fields:
$$f_*(u\cdot v)=f_*(u)\cdot f_*(v)\eqno(1.13)$$
(here $f_*$ is the induced linear map $f_*:T_xM\rightarrow T_{f(x)}M$).

{\bf Proposition 1.2.} {\it Algebraic symmetries of a Fr\"obenius manifold
form a finite-dimen\-si\-o\-nal Lie group $G(M)$.}

The generators of action of $G(M)$ on $M$ (i.e. the representation of
the Lie algebra of $G(M)$ in the Lie algebra of vector fields on
$M$) are the vector fields $w$ such that
$$[w,u\cdot v]=[w,u]\cdot v+[w,v]\cdot u\eqno(1.14)$$
for any vector fields $u$, $v$.

Note that the group $G(M)$ always is nontrivial: it contains the one-parameter
subgroup of shifts along the coordinate $t^1$. The generator of this subgroup
coincides with the unity vector field $e$.

The group $G(M)$ can be calculated for the important class of {\it massive}
Fr\"obenius manifolds.

{\bf Definition 1.4.} A Fr\"obenius manifold is called massive if the algebra
on $T_xM$ is semisimple for any $x\in M$.

In physical language massive Fr\"obenius manifolds are coupling spaces of
massive TFT models.

{\bf Main lemma.} {\it The connect component of the identity in the group
$G(M)$
of algebraic symmetries of a $n$-dimensional massive Fr\"obenius manifold is
a $n$-dimensional commutative Lie group that acts localy transitively on $M$.}

This is a reformulation of the main lemma of [39].

Action of the group of algebraic symmetries provides a new affine structure
on a massive Fr\"obenius manifold. The structure tensor $c_{ij}^k$ is constant
in this affine structure.

{}From the main lemma the following statement follows.

{\bf Theorem 1.1.} [39] {\it On a massive Fr\"obenius manifold local
coordinates
$u^1$,...$u^n$ exist such that the multiplication law of vector fields
in these coordinates has the form
$$\partial_i\cdot\partial_j=\delta_{ij}\partial_i,\eqno(1.15)$$
where $\partial_i=\partial /\partial u^i$. The invariant metric $\eta$ in these
coordinates has a diagonal form
$$\eta_{\alpha\beta}dt^\alpha dt^\beta =
\sum_{i=1}^n\eta_{ii}(u)(du^i)^2\eqno(1.16)$$
satisfying the equations
$$d(\sum_{i=1}^n\eta_{ii}(u)du^i)=0,\eqno(1.17a)$$
$$\sum_{k=1}^n\partial_k\eta_{ii}=0.\eqno(1.17b)$$
Conversely, for a flat diagonal metric with the properties (1.17)
and $t^\alpha = t^\alpha (u)$, $\alpha = 1,...,n$ being the flat coordinates
for the metric  the formulae
$$\eta_{\alpha\beta}=\sum_{i=1}^n\eta_{ii}(u){\partial u^i\over \partial
t^\alpha}{\partial u^i \over \partial t^\beta},\eqno(1.18a)$$
$$c_{\alpha\beta}^\gamma (t)=\sum_{i=1}^n{\partial u^i\over\partial t^\alpha}
{\partial u^i\over\partial t^\beta}{\partial t^\gamma\over\partial u^i},
\eqno(1.18b)$$
$$e^\alpha = \sum_{i=1}^n{\partial t^\alpha \over \partial u^i}\eqno(1.18c)$$
determine (localy) a massive Fr\"obenius manifold.}

The above coordinates $u^1$, ... , $u^n$ on a massive Fr\"obenius
manifold are determined uniquely up to permutations and shifts.
They are called {\it canonical coordinates} on the massive Fr\"obenius
manifold $M$. The tensor $c$ of structure constants in these coordinates
has the following canonical constant form
$$c_{ij}^k=\delta_{ij}\delta_j^k.\eqno(1.19)$$
The canonical coordinates $u^i$ can be found by solving an overdetermined
system of differential equations
$${\partial t^\alpha\over\partial u^i}{\partial t^\beta\over\partial
u^j}c_{\alpha\beta}^\gamma = \delta_{ij}{\partial t^\gamma\over\partial
u^i}.$$
For massive conformal invariant Fr\"obenius manifolds (see the next section)
they can be found in a pure algebraic way (the Proposition 2.4 below).

To complete local classification of massive TFT one has to classify flat
diagonal metrics with the properties (1.17). This class of metrics was
studied by Darboux [33] and Egoroff. Following Darboux, I will call
them {\it Egoroff metrics}. Vanishing of the curvature of these metrics
can be written in the form of the following system of PDE ({\it
Darboux -- Egoroff system}) for the
{\it rotation coefficients}
$$\gamma_{ij}(u)={\partial_j\sqrt{\eta_{ii}(u)}\over\sqrt{\eta_{jj}(u)}},
{}~i\neq j\eqno(1.20)$$
$$\partial_k\gamma_{ij}=\gamma_{ik}\gamma_{kj},
{}~~i,~j,~k ~{\rm are~ distinct},\eqno(1.21a)$$
$$\sum_{k=1}^n\partial_k\gamma_{ij}=0,~i\neq j\eqno(1.21b)$$
$$\gamma_{ji} = \gamma_{ij}.\eqno(1.21c)$$
It is interesting that the same equations (for even $n$) arise in the
calculation [34]
of multipoint correlators in impenetrable Bose-gas, see Appendix to [45].

Integrability of the Darboux -- Egoroff system was observed in [23].
It essentialy coincides with the \lq\lq pure imaginary reduction"
of the $n$-wave system (see [26, 25]). This
can be represented as the compatibility conditions of the following
linear system (depending on a spectral parameter $\lambda$)
$$\partial_j\psi_i=\gamma_{ij}\psi_j,~~i\neq j\eqno(1.22a)$$
$$\sum_{k=1}^n\partial_k\psi_i = \lambda \psi_i.\eqno(1.22b)$$

To complete local classification of massive Fr\"obenius manifolds one
first should apply an appropriate version of inverse spectral transform
(IST) to solve the Darboux -- Egoroff system (1.21). Below I will give
an example of IST for the important case of self-similar solutions of (1.21)
(so called topological conformal field theories [8, 51]). To find the metric
(1.16) and flat coordinates $t^\alpha = t^\alpha (u)$ for a given
solution $\gamma_{ij}(u)$ one has to fix a basis $\psi_{i\alpha}(u)$,
$\alpha = 1, ... n$ in the space of solutions of the system (1.22) for
$\lambda =0$
$$\partial_j\psi_{i\alpha} = \gamma_{ij}\psi_{j\alpha},
{}~~i\neq j,\eqno(1.23a)$$
$$\sum_k\partial_k\psi_{i\alpha} = 0,\eqno(1.23b)$$
$\alpha = 1,...,n.$
Then we put
$$\eta_{ii}(u) = \psi_{i1}^2(u),\eqno(1.24a)$$
$$\eta_{\alpha\beta} = \sum_{i=1}^n\psi_{i\alpha}(u)
\psi_{i\beta}(u),\eqno(1.24b)$$
$${\partial t_\alpha\over\partial u^i} = \psi_{i1}(u)\psi_{i\alpha}(u),
\eqno(1.24c)$$
$$c_{\alpha\beta\gamma}(t(u)) = \sum_{i=1}^n{\psi_{i\alpha}\psi_{i\beta}
\psi_{i\gamma}\over\psi_{i1}}.\eqno(1.24d)$$
These formulae complete local classification of complex massive Fr\"obenius
manifolds. They are parametrized (localy) by $n$ arbitrary functions
of 1 variable (the parametrization of solutions of the
Darboux -- Egoroff system) and also by $n$ complex parameters because of the
ambiguity in the choice of solutions $\psi_{i1}$ in the formulae (1.24).

To classify real Fr\"obenius manifolds one should apply IST to various
real forms of the Darboux -- Egoroff system. We will not do it here
(see [27] for discussion of real
forms of the system (1.21) in algebraic-geometry
IST).

Global topology of massive Fr\"obenius manifolds is rather poor. We say
that a $n$-dimensional manifold $M$ admits $S_n$-structure if the structure
group of the tangent bundle $TM$ can be reduced to the symmetric group
$S_n$. An atlas of coordinates charts on $M$ is {\it compatible} with the
given $S_n$-structure if differentials of the transition functions are
the correspondent elements of $S_n$ (in the standard $n$-dimensional
representation). Globaly a $S_n$-manifold $M$ with a compatible atlas
is determined by an affine representation of
$\pi_1(M)\rightarrow S_n\rightarrow Aff_n$, i.e. the transition functions
have the form
$$u^i\mapsto u^{\sigma (i)}+a_\sigma^i,\eqno(1.25a)$$
$$a_{\sigma '\sigma}^i = a_\sigma^{\sigma '(i)}
+a_{\sigma '}^i,\eqno(1.25b)$$
for $\sigma, ~\sigma '\in S_n$. As an example of $S_n$-manifold one can
have in mind the space of all polynomials
$M=\{P(u)=u^n+a_1u^{n-1}+...+a_n |~ a_1,...,a_n\in {\bf C}\}$
without multiple roots. Compatible coordinates are the roots of $P(u)$.
The transition functions (1.25) are given by the standard $n$-dimensional
representation of the braid group $\pi_1(M)=B_n$.

The Darboux -- Egoroff system is well-defined on any $S_n$-manifold
$M$ with a marked compatible atlas. To obtain a massive Fr\"obenius
structure on $M$ one should find a solution $\gamma_{ij}(u)$
being covariant with respect to transformations of the form (1.25).
This \lq\lq boundary value problem" seems to be more complicated.

In all the examples (below) of massive TFT the coupling space $M$
(massive Fr\"obenius manifold) can be extended by adding certain locus
$M_{sing}$ (at least of real codimension 2). The structure of
Fr\"obenius manifold can be extended on $\hat M = M\cup M_{sing}$
but the algebra structure on the tangent spaces $T_x\hat M$ for $x
\in M_{sing}$ has nilpotents. The flat metric $\eta_{\alpha\beta}$
is extended on $\hat M$ without degeneration. So $\hat M$ is still
a localy Euclidean manifold.

{\bf Remark.} The notion of Fr\"obenius manifolfd admits algebraic
formalization in terms of the ring of functions on a manifold.
More precisely, let $R$ be a commutative associative algebra
with a unity over a field $k$ of characteristics $\neq 2$. We
are interesting in structures of Fr\"obenius algebra over $R$
in the $R$-module of $k$-derivations $Der(R)$ (i.e. $u(\kappa
)
=0$ for $\kappa \in k, u\in Der(R)$) satisfying

$$\tilde\nabla_u(\lambda )\tilde\nabla_v(\lambda )-
\tilde\nabla_v(\lambda )\tilde\nabla_u(\lambda )=
\tilde\nabla_{[u,v]}(\lambda )~ ~{\rm
identicaly~in}~\lambda\eqno(1.26a)$$
$${\rm for}~\tilde\nabla_u(\lambda )v=\nabla_uv+\lambda u\cdot
v,\eqno(1.26b)$$
$$\nabla_ue=0~{\rm for ~all}~u\in Der(R)\eqno(1.26c)$$
where $e$ is the unity of the Fr\"obenius algebra $Der(R)$.
Non-degenerateness of the symmetric inner product
$$<~,~>:Der(R)\times Der(R)\to R$$
means that it provides an isomorphism ${\rm
Hom}_R(Der(R),R)\to Der(R)$. I recall that the covariant
derivative is a derivation $\nabla_uv\in Der(R)$ defined for
any $u,~v\in Der(R)$ being determined from the equation
$$<\nabla_uv,w>=$$
$${1\over
2}[u<v,w>+v<w,u>-w<u,v>+<[u,v],w>+<[w,u],v>+<[w,v],u>]
\eqno(1.27)$$
for any $w\in Der(R)$ (here $[~,~]$ denotes the commutator of
derivations). Note that the notion of infinitesimal algebraic
symmetry also can be algebraicaly
formalized in a similar way. It would be interesting to find a
pure algebraic version of the theorem 1.1  . This could give an
algebraic approach to the problem of classification of
Fr\"obenius manifolds.

We consider in conclusion of this section a closure of the class of massive
Fr\"obenius manifolds as the set of all Fr\"obenius manifolds with
$n$-dimensional commutative group of algebraic symmetries. Let $A$ be a fixed
$n$-dimensional Fr\"obenius
algebra with structure constants $c_{ij}^k$ and an invariant
inner nondegenerate inner product $\epsilon = (\epsilon_{ij})$. Let us
introduce matrices
$$C_i = (c_{ij}^k).\eqno(1.28)$$
An analogue of the Darboux -- Egoroff system (1.21) for an operator-valued
function
$$\gamma (u):A\to A,~\gamma =(\gamma_i^j(u)),~u=(u^1,\dots ,u^n)\eqno(1.29a)$$
(an analogue of the rotation coefficients) where the operator $\gamma$ is
symmetric with respect to $\epsilon$,
$$\epsilon\gamma = \gamma^{{\rm T}}\epsilon\eqno(1.29b)$$
has the form
$$[C_i,\partial_j\gamma ]-[C_j,\partial_i\gamma ]+
[[C_i,\gamma ],[C_j,\gamma ]] = 0,~i,j=1,\dots ,n,\eqno(1.30)$$
$\partial_i=\partial /\partial u^i$. This is an integrable system with the
Lax representation
$$\partial_i\Psi = \Psi (\lambda C_i +[C_i,\gamma ]),~i=1,\dots
,n.\eqno(1.31)$$
It is convenient to consider $\Psi = (\psi_1(u),\dots ,\psi_n(u))$ as a
function with values in the dual space $A^*$. Note that $A^*$ also is a
Fr\"obenius algebra with the structure constants $c^{ij}_k =
c_{ks}^i\epsilon^{sj}$ and the invariant inner product $<~,~>_*$ determined
by $(\epsilon^{ij}) = (\epsilon_{ij})^{-1}$.

Let $\Psi_\alpha (u)$, $\alpha = 1$, ... ,$n$ be a basis of solutions of
(1.31) for $\lambda =0$
$$\partial \Psi_\alpha = \Psi_\alpha [C_i,\gamma ],~
\alpha =1,... ,n\eqno (1.32a)$$
such that the vector $\Psi_1(u)$ is invertible in $A^*$. We put
$$\eta_{\alpha\beta} = <\Psi_\alpha (u),\Psi_\beta (u)>_*\eqno(1.32b)$$
$${\rm grad}_u t_\alpha  = \Psi_\alpha (u)\cdot
\Psi_1 (u)\eqno(1.32c)$$
$$c_{\alpha\beta\gamma}(t(u)) =
{\Psi_\alpha (u)\cdot\Psi_\beta (u)\cdot\Psi_\gamma (u)
\over \Psi_1(u)}.\eqno(1.32d)$$

{\bf Theorem 1.2.} {\it Formulae (1.32) for arbitrary Fr\"obenius
algebra $A$ localy parametrize all Fr\"obenius manifolds with
$n$-dimensional commutative group of algebraic symmetries.}

Considering $u$ as a vector in $A$ and $\Psi_1^2=\Psi_1\cdot\Psi_1$ as
a linear function on $A$ one obtains the following analogue of Egoroff
metrics (on $A$)
$$ds^2 = \Psi_1^2(du\cdot du).\eqno(1.33)$$
\medskip
{\bf 2. Conformal invariant Fr\"obenius manifolds and isomonodromy
deformations.}
\medskip
{\bf Definition 2.1.} A diffeomorphism $f:M\to M$ is called
{\it conformal symmetry} if
$$f_*(u\cdot v) = \mu_f^cf_*(u)\cdot f_*(v)\eqno(2.1a)$$
$$<f_*(u),f_*(v)> = \mu_f^\eta <u,v>\eqno(2.1b)$$
$$f_*(e) = \mu_f^ee\eqno(2.1c)$$
for some functions $\mu_f^c$, $\mu_f^\eta$, $\mu_f^e$.
A Fr\"obenius manifold $M$ is called {\it conformal invariant} if it admits
a one-parameter group of conformal symmetries $f^{(\tau )}$ such that the
tensors $f_*^{(\tau )}(c)$, $f_*^{(\tau )}(\eta )$, $f_*^{(\tau )}(e)$
determine on $M$ a Fr\"obenius structure for any $\tau$.

Let $v$ be the generator of the one-parameter group of conformal symmetries
on a conformal invariant Fr\"obenius manifolds.

{\bf Proposition 2.1.} {\it On a massive conformal invariant Fr\"obenius
manifold an
action of the one-parameter group of conformal symmetries is generated
by the field
$$v=\sum_{i=1}^nu^i\partial_i\eqno(2.2)$$
(modulo obvious transformations $v\mapsto av+be$ for constant
$a$ and $b$).
It acts on the tensors $c$, $\eta$, $e$ by the following formulae
$${\cal L}_vc = c\eqno(2.3a)$$
$${\cal L}_ve = -e\eqno(2.3b)$$
$${\cal L}_v\eta = (2-d)\eta\eqno(2.3c)$$
where $d$ is a constant.}

Here ${\cal L}_v$ means the Lie derivative along the vector field $v$.

{\bf Corollary.} {\it For a massive conformal invariant Fr\"obenius manifold
the rotation coefficients $\gamma_{ij}(u)$ satisfy the similarity
condition
$$\gamma_{ij}(cu) = c^{-1}\gamma_{ij}(u)\eqno(2.4a)$$
or, equivalently
$$\sum_{k=1}^nu^k\partial_k \gamma_{ij}(u) = -\gamma_{ij}(u).\eqno(2.4b)$$}

For $n=2$ the similarity reduction (2.4) of the Darboux -- Egoroff
system can be solved immediately:
$$\gamma_{12} = \gamma_{21} = {id\over 2}{1\over u^1-u^2}.\eqno(2.5)$$
For the first nontrivial case $n=3$ the system (1.21), (2.4) reads
$$\Gamma_1' = \Gamma_2\Gamma_3\eqno(2.6a)$$
$$(z\Gamma_2)' = -\Gamma_1\Gamma_3\eqno(2.6b)$$
$$((z-1)\Gamma_3)' = \Gamma_1\Gamma_2\eqno(2.6c)$$
where
$$\gamma_{ij}(u) = {1\over u^2-u^3}\Gamma_k(z), ~ i,~j,~k~{\rm are
{}~distinct}\eqno(2.7a)$$
$$z={u^1-u^3\over u^2-u^3}.\eqno(2.7b)$$
It has an obvious first integral
$$\Gamma_1^2+(z\Gamma_2)^2+((z-1)\Gamma_3)^2 = {\rm const}.\eqno(2.8)$$
Using this integral one can reduce [30] the system (2.6) to a particular case
of the Painlev\'e-VI equation.

For $n>3$ the system (1.21), (2.4) can be considered as a high-order
analogue of the Painlev\'e-VI. To find solutions of this system
one can use an appropriate version of IST: the so-called method
of isomonodromy deformations [31]. This gives parametrization of
solutions of the system (1.21), (2.4) by monodromy data of the following system
of linear ODE with rational coefficients:
$$\lambda{d\psi\over d\lambda} =
(\lambda U - [U,\gamma ])\psi .\eqno(2.9a)$$
Here
$$U = {\rm diag}(u^1,...,u^n),\eqno(2.9b)$$
$$\gamma = (\gamma_{ij}(u)).\eqno(2.9c)$$
Solutions of this linear ODE have some monodromy properties,
i.e. they are multivalued functions in the complex
$\lambda$-plane.

{\bf Proposition 2.2.} {\it The monodromy transformations of solutions
of the system (2.9) do not depend on the parameters $u$ iff
the matrix $\gamma_{ij}(u)$ is a solution of the system (1.21), (2.4).}

The linear system (2.9) has two singular points: a regular singularity
at $\lambda = 0$ and an irregular one at $\lambda = \infty$.
Monodromy transformations of solutions of the system near $\lambda = 0$
have the form
$$\psi\mapsto\exp (-2\pi i[U,\gamma ])\psi .\eqno(2.10)$$
So the eigenvalues of the matrix $[U,\gamma ]$ are first integrals
of the system (1.21), (2.4).
These generalise the first integral (2.8). Monodromy
at infinity is determined by a $n \times n$ {\it Stokes matrix} $S$
(see [31] for details). The diagonal terms of $S$ equal 1; $n(n-1)/2$
of other entries of the matrix $S$ vanish. Other matrix elements
of $S$ can be used as local parameters of massive conformal-invariant
Fr\"obenius manifolds (just $n(n-1)/2$ arbitrary complex
parameters;
one should add one more parameter: a norming constant of a solution
$\psi_{i1}(u)$ in (1.24) being an eigenvector of the matrix
$[U,\gamma ]$). The monodromy at $\lambda = 0$ can be expressed
via $S$ using cyclic relations (see [39]). If the Stokes matrix $S$ is
sufficiently close to the unity matrix then the inverse problem
of the monodromy theory (i.e., to determine the coefficients of
the linear operator (2.9) from the given monodromy data) always is
solvable. The solution can be obtained by solving linear integral
equations [39].

Let us assume that the monodromy of the operator (2.9) in the origin
is semisimple. That means that the matrix $[U,\gamma ]$ has pairwise
different eigenvalues $\mu_1$, ... $\mu_n$. Let us order them in such
a way that
$$\mu_\alpha +\mu_{n-\alpha +1}=0.\eqno(2.11)$$

{\bf Proposition 2.3.} {\it Flat coordinates on a massive conformal invariant
Fr\"obenius manifold with semisimple monodromy of (2.9) at $\lambda = 0$
can be chosen in such a way that the generator $v$ of conformal symmetries
has the form
$$v = \sum (1-q_\alpha )t^\alpha \partial_\alpha\eqno(2.12)$$
for
$$q_\alpha = \mu_1-\mu_\alpha\eqno(2.13a)$$
where $\mu_\alpha$ are the eigenvalues of the matrix $[U, \gamma ]$
ordered as in (2.11).}

In other words, the tensors $c$, $\eta$, $e$ should be conformal
covariant with respect to the following transformations
$$t^\alpha\mapsto c^{1-q_\alpha }t^\alpha\eqno(2.14a)$$
$$c_{\alpha\beta}^\gamma\mapsto c^{q_\alpha +q_\beta -q_\gamma}
c_{\alpha\beta}^\gamma\eqno(2.14b)$$
$$\eta_{\alpha\beta}\mapsto c^{q_\alpha +q_\beta -d}\eta_{\alpha\beta}
\eqno(2.14c)$$
where
$$d = q_n = 2\mu_1\eqno(2.13b)$$
is the same as in (2.3c),
$$e\mapsto c^{-1}e.\eqno(2.14d)$$
The equation (2.14c) means that $\eta_{\alpha\beta}\neq 0$ only for
$q_\alpha +q_\beta = d$.

The numbers $q_\alpha$ are called {\it charges} of the TCFT model,
$d$ is called {\it dimension} of the model. For topological
sigma-models it coincides with complex dimension of the target-space.
Scaling laws (2.14)
were obtained in [8] using the assumption that the TCFT model is
obtained by twisting of a N=2 supersymmetric model of QFT. These
imply superselection rules for tree-level correlators in the
conformal point $t=0$ (the stationary point of the field $v$ (2.3)).
In our approach the scaling laws follow from simple symmetry
assumption on the Fr\"obenius manifold.

Summarizing we obtain

{\bf Theorem 2.1.} {\it All massive conformal invariant Fr\"obenius
manifolds are parametrized by monodromy data of the linear operator
$$\Lambda =\lambda \partial_\lambda - \lambda U + M(u)\eqno(2.15a)$$
$$U={\rm diag}(u^1,\dots ,u^n).\eqno(2.15b)$$
$$M^{{\rm T}} = -M.\eqno(2.15c)$$
Manifolds with semisimple monodromy at the origin $\lambda = 0$ form a
$[{n(n-1)\over 2}+1]$-parameter family. The free energy $F(t)$ of such a
Fr\"obenius manifold can be expressed via quadratures of a high-order analogue
of the Painlev\'e-VI transcendents, i.e. solutions of the equations of
isomonodromy deformations of (2.15).}

For  nonresonant conformal invariant Fr\"obenius manifolds (see (3.19)
below) with a semisimple monodromy at the origin the structure functions
$c_{\alpha\beta}^\gamma (t)$ can be expressed algebraicaly (i.e. without
quadratures) via the above high-order analogue of the Painlev\'e-VI
transcendents. Also one has

{\bf Proposition 2.4.} {\it The canonical coordinates $u^1,\dots ,u^n$
on a massive conformal invariant Fr\"obenius manifold coincide with
eigenvalues of the matrix
$$\tilde U = (\tilde U_\beta^\gamma (t)) =
((1+q_\beta -q_\gamma )F_\beta^\gamma (t))\eqno(2.16a)$$
$$F_\beta^\gamma (t) = \eta^{\gamma\epsilon}\partial_\beta
\partial_\epsilon F(t).\eqno(2.16b)$$}

It would be interesting to understand a physical sense of the
operator $\tilde U$ for TCFT models.

{\bf Remark.} We saw that monodromy is an important invariant of a massive
conformal invariant Fr\"obenius manifold. It can be defined also for arbitrary
conformal invariant Fr\"obenius manifold by considering the linear operator
$$\tilde\Lambda = \lambda\partial_\lambda -\lambda \tilde U +\tilde M,
\eqno(2,17a)$$
where
$$\tilde M = (\tilde M_\beta^\gamma ) =(q_\beta\delta_\beta^\gamma ),
\eqno(2.17b)$$
the matrix $\tilde U$ has the form (2.16).
WDVV equations determine isomonodromy deformations of $\tilde\Lambda$.

Monodromy properties of eigenfunctions of $\tilde\Lambda$ near irregular
singularity $\lambda =\infty$ (i.e. Stokes matrices)
strongly depend on algebraic structure of
the multiplication on $TM$. These Stokes matrices are constrained by
cyclic relations since monodromy near the origin $\lambda =0$ is
fixed by the given charges $q_\alpha$. An advantage of the isomonodromy
problem (2.15) for massive Fr\"obenius manifolds is in universality
(independence on the charges; the charges can be expressed via an
arbitrary Stokes matrix of (2.15)). Note that a basis of common eigenfunctions
of $\tilde\Lambda$
$$\tilde\Lambda\xi = \kappa\xi\eqno(2.18a)$$
and (1.12) has the form
$$\xi_\beta (t,\lambda ) = \partial_\beta h_\alpha (t,\lambda ),
{}~\kappa = d-q_\alpha,~{\rm for~any~}\alpha =1,\dots ,n\eqno(2.18b)$$
where the solutions $h_\alpha (t,\lambda )$ of (3.5) are normalized by (3.6).
 \medskip
{\bf 3. Coupling to gravity. Systems of hydrodynamic type: their
Hamiltonian formalism, solutions, and $\tau$-functions.}
\medskip
Let us fix a Fr\"obenius manifold (i.e. a solution of the WDVV equations.
Considering this as the primary free energy of the
matter sector of a 2D TFT model, let us try to calculate the
tree-level (i.e., the zero-genus) approximation of the complete model
obtained by coupling of the matter sector to topological gravity. The
idea to use hierarchies of  Hamiltonian systems of hydrodynamic type for such a
calculation was proposed by E.Witten [46] for the case of topological
sigma-models. An advantage of my approach is in effective construction
of these hierarchies for any solution of WDVV. The tree-level free
energy of the model will be identified with $\tau$-function of a
particular solution of the hierarchy. For a TCFT-model (i.e. for a
conformal invariant Fr\"obenius manifold) the hierarchy carries a
bihamiltonian structure under a non-resonance assumption for charges
and dimension of the model (this bihamiltonian structure was
constructed in [39] for the case of massive perturbations of a TCFT
model; here I generalize it for an arbitrary TCFT model).
This gives an answer to a question of [46] (see p.283). As it was
mentioned in the Introduction, the bihamiltonian structure could be
useful for calculation higher genus corrections .

So let $c_{\alpha\beta}^\gamma (t)$, $\eta_{\alpha\beta}$ be a solution of
WDVV, $t=(t^1,\dots , t^n)$. I will construct a hierarchy of the first
order PDE systems linear in derivatives ({\it systems of hydrodynamic
type}) for functions $t^\alpha (T)$, $T$ is an infinite vector
$$T=(T^{\alpha ,p}),~ ~\alpha =1,~\dots ,~n,~~p=0,~1,~\dots ;
{}~T^{1,0}=X,$$
$$\partial_{T^{\alpha ,p}}t^\beta = {c_{(\alpha ,p)}}_\gamma^\beta (t)
\partial_X t^\gamma \eqno(3.1a)$$
for some matrices of coefficients
${c_{(\alpha ,p)}}_\gamma^\beta (t)$.
The marked variable $X=T^{1,0}$ usualy is called {\it cosmological
constant}.

I will consider the equations (3.1) as dynamical systems (for any
$(\alpha , p)$) on the space of functions $t=t(X)$ with values in the
Fr\"obenius manifold $M$.

A. Construction of the systems. I define a Poisson bracket on the
space of functions $t=t(X)$ (i.e. on the loop space ${\cal L}(M)$) by the
formula
$$\{ t^\alpha (X),t^\beta (Y)\} = \eta^{\alpha\beta}\delta '(X-Y).
\eqno(3.2)$$
All the systems (3.1a) have hamiltonian form
$$\partial_{T^{\alpha ,p}}t^\beta = \{ t^\beta (X), H_{\alpha ,
p}\} \eqno(3.1b)$$
with the Hamiltonians of the form
$$H_{\alpha , p} = \int h_{\alpha , p+1}(t(X))dX. \eqno (3.3)$$
The generating functions of densities of the Hamiltonians
$$h_\alpha (t,\lambda ) = \sum_{p=0}^\infty h_{\alpha , p}(t)
\lambda^p,~\alpha =1,\dots ,n \eqno (3.4)$$
coincide with the flat coordinates of the perturbed connection
$\tilde\nabla (\lambda )$ (see (1.10)). That means that they are
determined by the system (cf. (1.12))
$$\partial_\beta\partial_\gamma h_\alpha (t,\lambda ) = \lambda
c_{\beta\gamma}^\epsilon (t) \partial_\epsilon h_\alpha (t,\lambda ).
\eqno(3.5)$$
This gives simple recurrence relations for the densities $h_{\alpha
,p}$. Solutions of (3.5) can be normalized in such a way that
$$h_\alpha (t,0) = t_\alpha =\eta_{\alpha\beta}t^\beta ,\eqno(3.6a)$$
$$<\nabla h_\alpha (t,\lambda ),\nabla h_\beta (t,-\lambda )> =
\eta_{\alpha\beta}. \eqno(3.6b)$$
Here $\nabla$ is the gradient (in $t$).
It can be shown that the Hamiltonians (3.3) are in involution. So all
the systems of the hierarchy (3.1) commute pairwise.

B. Specification of a solution $t=t(T)$. The hierarchy (3.1) admits an
obvious scaling group
$$T^{\alpha ,p} \mapsto cT^{\alpha ,p},~~t\mapsto t.\eqno(3.7)$$
Let us take the nonconstant invariant solution for the symmetry
$$(\partial_{T^{1,1}} - \sum T^{\alpha ,p}\partial_{T^{\alpha ,p}})
t(T) = 0 \eqno(3.8)$$
(I identify $T^{1,0}$ and $X$. So the variable $X$ is supressed in the
formulae.) This solution can be found without quadratures from a fixed
point equation for the gradient map
$$t=\nabla \Phi_T(t),\eqno(3.9)$$
$$\Phi_T(t) = \sum_{\alpha ,p}T^{\alpha ,p}h_{\alpha ,p}(t).\eqno(3.10)$$
It can be proved existence and uniqueness of such a fixed point for
sufficiently small $T^{\alpha ,p}$ for $p>0$ (more precisely, in the
domain: $T^{\alpha ,0} $ are arbitrary, $T^{1,1} = o(1)$, $T^{\alpha
,p} = o(T^{1,1})$ for $p>0$).

C. $\tau$-function. Let us define coefficients $V_{(\alpha ,p),(\beta
,q)}(t)$ from the expansion
$$(\lambda +\mu )^{-1}(<\nabla h_\alpha (t,\lambda ), \nabla h_\beta
(t,\mu)> - \eta_{\alpha\beta}) = \sum_{p,q=0}^\infty V_{(\alpha ,p),
(\beta ,q)}(t)\lambda^p\mu^q \equiv V_{\alpha\beta}(t,\lambda ,\mu ).
\eqno(3.11)$$
The infinite matrix of coefficients $V_{(\alpha ,p),(\beta ,q)}(t)$ has a
simple meaning: it is the energy-momentum tensor of the commutative
Hamiltonian hierarchy (3.1). That means that
a matrix entry $V_{(\alpha ,p),(\beta ,q)}(t)$
is the density of flux of the Hamiltonian $H_{\alpha ,p}$ along the flow
$T^{\beta ,q}$:
$$\partial_{T^{\beta ,q}}h_{\alpha ,p+1}(t) = \partial_X
V_{(\alpha ,p),(\beta ,q)}(t).\eqno(3.12)$$
Then
$$\tau (T) = {1\over 2}\sum V_{(\alpha ,p),(\beta ,q)}(t(T))
T^{\alpha, p}T^{\beta ,q} + \sum V_{(\alpha ,p),(1,1)}(t(T)) T^{\alpha
,p} + {1\over 2}V_{(1,1),(1,1)} (t(T))\eqno(3.13)$$

{\bf Remark.} More general family of solutions of (3.1) has the form
$$\nabla [\Phi_T(t)-\Phi_{T_0}(t)] = 0 \eqno(3.14)$$
for arbitrary constant vector $T_0 = T_0^{\alpha ,p}$. For massive Fr\"obenius
manifolds these form a dense subset in the space of all solutions of (3.1)
(see [22, 23, 39]). Formally
they can be obtained from the solution (3.9) by a shift of
the arguments $T^{\alpha ,p}$.
$\tau$-function of the solution (3.14) can be formaly obtained from
(3.13) by the same shift. For the example of topological gravity [3, 46]
such a shift is just the operation that relates the tree-level free energies
of the topological phase of 2D gravity and of the matrix model. It should be
taken in account that the operation of such a time shift in systems of
hydrodynamic type is a subtle one: it can pass through a point of gradient
catastrophe where derivatives become infinite. The correspondent solution of
the KdV hierarchy has no gradient catastrophes but oscillating zones arise
(see [32] for details).

{\bf Theorem 3.1}. {\it Let
$${\cal F}(T)=\log \tau (T),\eqno(3.15a)$$
$$<\phi_{\alpha ,p} \phi_{\beta ,q}\dots >_0 = \partial_{T^{\alpha
,p}} \partial_{T^{\beta ,q}}\dots {\cal F}(T).\eqno(3.15b)$$
Then the following relations hold
$${\cal F}(T)|_{_{T^{\alpha ,p}=0 ~{\rm for}~p>0,~T^{\alpha
,0}=t^\alpha}} = F(t) \eqno(3.16a)$$
$$\partial_X{\cal F}(T) = \sum T^{\alpha ,p}\partial_{T^{\alpha ,p-1}}
{\cal F}(T)+ {1\over 2}
\eta_{\alpha\beta}T^{\alpha ,0}T^{\beta ,0} \eqno(3.16b)$$
$$<\phi_{\alpha ,p}\phi_{\beta ,q}\phi_{\gamma ,r}>_0 = <\phi_{\alpha
,p-1}\phi_{\lambda ,0}>_0 \eta^{\lambda\mu} <\phi_{\mu ,0} \phi_{\beta
,q} \phi_{\gamma ,r}>_0.\eqno(3.16c)$$}

Let me establish now a 1-1 correspondence between the statements of
the theorem and the standard terminology of QFT. In a complete model
of 2D TFT (i.e. a matter sector coupled to topological gravity) there
are infinite number of operators. They usualy are denoted by
$\phi_{\alpha ,p}$ or $\sigma_p(\phi_\alpha )$. The operators
$\phi_{\alpha ,0}$ can be identified with the primary operators
$\phi_\alpha$; the operators $\phi_{\alpha ,p}$ for $p>0$ are called
{\it gravitational descendants} of $\phi_\alpha$. Respectively one has
infinite number of coupling constants $T^{\alpha ,p}$. The formula (3.15a)
expresses the tree-level (i.e. genus zero)
partition function of the model of 2D TFT
via logarythm of
the $\tau$-function (3.13). Equation (3.15b) is the standard relation between
the correlators (of genus zero) in the model and the free energy.
Equation (3.16a) manifests that before coupling to gravity the partition
function (3.15a) coincides with the primary partition function of the
given matter sector. Equation (3.16b) is the string equation for the free
energy [3, 4, 8, 46].
And equations (3.16c) coincide with the genus zero recursion
relations for correlators of a TFT [4, 46].

Particularly, from (3.15) one obtains
$$<\phi_{\alpha ,p}\phi_{\beta ,q}>_0 = V_{(\alpha ,p),(\beta ,q)}
(t(T)),\eqno(3.17a)$$
$$<\phi_{\alpha ,p}\phi_{1,0}>_0 =h_{\alpha ,p}(t(T)),\eqno(3.17b)$$
$$<\phi_{\alpha ,p}\phi_{\beta ,q}\phi_{\gamma ,r}>_0 =
<\nabla h_{\alpha ,p}\cdot\nabla h_{\beta ,q}\cdot\nabla h_{\gamma
,r},[e-\sum T^{\alpha ,p}\nabla h_{\alpha ,p-1}]^{-1}>.\eqno(3.17c)$$
The second factor of the inner product in the r.h.s. of (3.17c) is
an invertible element (in the Fr\"obenius algebra of vector fields on $M$) for
sufficiently small $T^{\alpha ,p}$, $p>0$. From the last formula one obtains

{\bf Proposition 3.1.} {\it The coefficients
$$c_p,_{\alpha\beta}^\gamma (T) = \eta^{\gamma\mu}
\partial_{T^{\alpha ,p}}
\partial_{T^{\beta ,p}}
\partial_{T^{\mu ,p}}\log \tau (T) \eqno(3.18)$$
for any $p$ and any $T$ are structure constants of a commutative
associative algebra with the invariant inner product $\eta_{\alpha\beta}$.}

As a rule such an algebra has no unity.

In fact the Proposition holds also for a $\tau$-function of an arbitrary
solution of the form (3.14).

We see that the hierarchy (3.1) determines a family of B\"acklund
transforms of the WDVV equation (0.1)
$$F(t)\mapsto \tilde F(\tilde t),$$
$$\tilde F = \log\tau ,~\tilde t^\alpha = T^{\alpha ,p}$$
for a fixed $p$ and for arbitrary $\tau$-function of (3.1). So it is
natural to consider equations of the hierarchy as Lie -- B\"acklund
symmetries of WDVV.

Up to now I even did not use the scaling invariance (2.14). It turns out
that this gives rise to a bihamiltonian structure of the hierarchy
(3.1).

Let us consider a conformal invariant Fr\"obenius manifold, i.e. a
TCFT model with charges $q_\alpha$ and dimension $d$.
We say that a pair $\alpha ,p$ is {\it resonant} if
$${d+1 \over 2} -q_\alpha +p = 0.\eqno(3.19)$$
Here $p$ is a nonnegative integer. The TCFT model is {\it nonresonant}
if all pairs $\alpha ,p$ are nonresonant. For example, models
satisfying the inequalities
$$0=q_1\leq q_2\leq\dots\leq q_n=d<1 \eqno(3.20)$$
all are nonresonant.

{\bf Theorem 3.2}. {\it 1) For a conformal invariant Fr\"obenius manifold
 with charges $q_\alpha$ and
dimension $d$ the formula
$$\{t^\alpha (X), t^\beta (Y)\}_1 =[({d+1\over 2}-q_\alpha )
F^{\alpha\beta}(t(X)) + ({d+1\over 2}-q_\beta ) F^{\alpha\beta}(t(Y))]
\delta '(X-Y) \eqno(3.21)$$
$$F^{\alpha\beta}(t) = \eta^{\alpha\alpha '}\eta^{\beta\beta '}
\partial_{\alpha '}\partial_{\beta '}F(t)$$
determines a Poisson bracket compatible with the Poisson bracket (3.2).
2) For a nonresonant TCFT model all the equations of the hierarchy (3.1)
are Hamiltonian equations also with respect to the Poisson bracket
(3.21).}

The nonresonancy condition is essential: equations (3.1) with resonant
numbers $(\alpha ,p)$ do not admit another Poisson structure.

{\bf Remark}. According to the theory [18-21] of Poisson brackets
of hydrodynamic type any such a bracket is determined by a
flat Riemannian (or pseudo-Riemannian) metric $g_{\alpha\beta}(t)$
on the target
space $M$ (more precisely, one needs a metric
$g^{\alpha\beta}(t)$ on the cotangent bundle to $M$).
In our case the target space is the Fr\"obenius
manifold $M$. The first Poisson structure (3.2) is determined by
the metric being specified by the double-point correlators
$\eta_{\alpha\beta}$. The second flat metric for the Poisson
bracket (3.21) on a conformal invariant Fr\"obenius manifold $M$
has the following geometrical interpretation. Let $\omega_1$
and $\omega_2$ be two 2-forms on $M$. We can multiply them
$\omega_1,~\omega_2\mapsto \omega_1\cdot\omega_2$ using the
multiplication of tangent vectors and the isomorphism $\eta$
between tangent and cotangent spaces. Then the new inner
product $<~,~>_1$ is defined by the formula
$$<\omega_1,\omega_2>_1 = {\rm i}_v(\omega_1\cdot\omega_2).
\eqno(3.22)$$
Here ${\rm i}_v$ is the operator of contraction with the
vector field $v$ (the generator of conformal symmetries (2.3)).
The metric (3.22) can be degenerate. The theorem states that,
nevertheless, the Jacobi identity for the Poisson bracket (3.21)
holds.

Main examples of solutions of WDVV and of corresponding
hierarchies will be given in the next section. Here I will
consider the simplest class of examples where
$c_{\alpha\beta}^\gamma$ does not depend on $t$. They form
structure constants of a Fr\"obenius algebra $A$ with an
invariant inner product $<~,~>$ ($\eta_{\alpha\beta}$ in a
basis $e_1=1,\dots ,e_n$). Let
$${\bf t}=t^\alpha e_\alpha \in A.\eqno(3.23)$$
The linear system (3.5) can be solved easily:
$$h_\alpha (t,\lambda )=\lambda^{-1}<e_\alpha , e^{\lambda {\bf
t}}
-1>.$$
This gives the following form of the hierarchy (3.1)
$$\partial_{T^{\alpha ,p}}{\bf t} = {1\over p!}e_\alpha {\bf
t}^p\partial_X{\bf t}.\eqno(3.24)$$
The solution (3.9) is specified as the fixed point
$$G({\bf t})={\bf t},\eqno(3.25a)$$
$$G({\bf t}) = \sum_{p=0}^\infty {{\bf T}_p\over p!}{\bf t}^p.
\eqno(3.25b)$$
Here I introduce $A$-valued coupling constants
$${\bf T}_p = T^{\alpha ,p}e_\alpha\in A,~p=0,~1,\dots .
\eqno(3.26)$$
The solution of (3.25) has the well-known form
$${\bf t} =G(G(G(\dots ))) \eqno(3.27)$$
(infinite number of iterations). The $\tau$-function of the
solution (3.27) has the form
$$\log\tau = {1\over 6}<1,{\bf t}^3> - \sum_p{<{\bf T}_p, {\bf
t}^{p+2}>\over (p+2)p!} + {1\over 2}\sum_{p,q} {<{\bf T}_p
{\bf T}_q,{\bf t}^{p+q+1}>\over (p+q+1)p!q!}. \eqno(3.28)$$
For the tree-level correlation functions of a TFT-model with constant
primary correlators one immediately obtains
$$<\phi_{\alpha ,p}\phi_{\beta ,q}>_0 = {<e_{\alpha}e_{\beta },
{\bf t}^{p+q+1}>\over (p+q+1)p!q!}, \eqno(3.29a)$$
$$<\phi_{\alpha ,p}\phi_{\beta ,q}\phi_{\gamma ,r}>_0 =
{1\over p!q!r!}<e_\alpha e_\beta e_\gamma ,{{\bf
t}^{p+q+r}\over 1 - \sum_{s\geq 1}{{\bf T}_s{\bf t}^{s-1}
\over (s-1)!}}>. \eqno(3.29b)$$
For $n=1$ the formulae (3.29) give the tree-level
correlators of the topological gravity (see [3, 46]). For $n=24$
one obtains the tree-level correlators of the topological
sigma-model with a K3-surface as the target space. Here the
algebra $A=H^*(K3)$ is a graded one: it has a basis $P$, $Q_1$, ...,
$Q_{22}$, $R$ of degrees 0, 1 (all the $Q$'s) and 2 resp. The
multiplication has the form
$$P~{\rm is~the~unity},~Q_iQ_j =\eta_{ij}R,~ Q_iR=R^2=0
\eqno(3.30)$$
for a nondegenerate symmetric matrix $\eta_{ij}$. The scalar
product (the intersection number) has the form
$$\eta_{PR} =1, ~\eta_{Q_iQ_j} = \eta_{ij}.$$

Let us consider now the second hamiltonian structure (3.21). I
start with the most elementary case $n=1$ (the pure gravity).
Let me redenote the coupling constant
$$u=t^1.$$
The Poisson bracket (3.21) for this case reads
$$\{ u(X),u(Y)\}_1 ={1\over 2}(u(X)+u(Y))\delta '(X-Y).
\eqno(3.31)$$
This is nothing but the Lie -- Poisson bracket on the dual
space to the Lie algebra of one-dimensional vector fields.

For arbitrary graded Fr\"obenius algebra $A$ the Poisson
bracket (3.21) also is linear in the coordinates $t^\alpha$
$$\{ t^\alpha (X),t^\beta (Y)\}_1 = [({d+1\over 2} - q_\alpha
)c^{\alpha\beta}_\gamma t^\gamma (X) + ({d+1\over 2} - q_\beta
)c^{\alpha\beta}_\gamma t^\gamma (Y)]\delta '(X-Y). \eqno(3.32)$$
It determines therefore a structure of an infinite dimensional
Lie algebra on the loop space ${\cal L}(A^*)$ where $A^*$ is
the dual space to the graded Fr\"obenius algebra $A$. Theory
of linear Poisson brackets of hydrodynamic type and of
corresponding infinite dimensional Lie algebras was
constructed in [34] (see also [18]). But the class of examples
(3.32) is a new one. Note that the case $A=H^*(K3)$ is a nonresonant one.

Let us come back to the general (i.e. nonlinear) case of a
TCFT model. I will assume that the charges and the dimension
are ordered in such a way that
$$0=q_1< q_2\leq\dots\leq q_{n-1}<q_n=d.\eqno(3.33)$$
Then from (3.21) one obtains
$$\{ t^n(X),t^n(Y)\}_1 = {1-d\over 2}(t^n(X)+t^n(Y))\delta
'(X-Y). \eqno(3.34)$$
Since
$$\{ t^\alpha (X), t^n(Y)\}_1 = [({d+1\over 2}-q_\alpha )t^\alpha
(X) +{1-d\over 2}t^\alpha (Y)]\delta '(X-Y), \eqno(3.35)$$
the functional
$$P = {2\over 1-d}\int t^n(X)dX \eqno(3.36)$$
generates spatial translations.
We see that for $d\neq 1$ the Poisson bracket (3.21) can be
considered as a nonlinear extension of the Lie algebra of
one-dimensional vector fields. An interesting question is to
find an analogue of the Gelfand -- Fuchs cocycle for this
bracket. I found such a cocycle for a more particular class of
TCFT models. We say that a TCFT-model is {\it graded} if for
any $t$ the Fr\"obenius algebra $c_{\alpha\beta}^\gamma (t)$,
$\eta_{\alpha\beta}$ is graded.

{\bf Theorem 3.3}. {\it For a graded TCFT-model the formula
$$\{ t^\alpha (X),t^\beta (Y) \}^{\hat{}} _1 =
\{ t^\alpha (X),t^\beta (Y) \} _1 + \epsilon^2\eta^{1\alpha}\eta^{1\beta}
\delta '''(X-Y) \eqno(3.37)$$
determines a Poisson bracket compatible with (3.2) and (3.21) for
arbitrary $\epsilon^2$ (the central charge). For a generic graded TCFT
model this is the only one deformation of the Poisson bracket
(3.21) proportional to $\delta '''(X-Y)$.}

For $n=1$ (3.37) determines nothing but the Lie -- Poisson
bracket on the dual  space to the Virasoro algebra
$$\{ u(X),u(Y)\} _1^{\hat{}}  = {1\over 2}[u(X)+u(Y)] \delta '(X-Y) +\epsilon^2
\delta '''(X-Y)\eqno(3.38)$$
(the second Poisson structure of the KdV hierarchy).
For $n>1$ and constant primary correlators (i.e. for a
constant graded Fr\"obenius algebra $A$) the Poisson bracket
(3.37) can be considered as a vector-valued extension (for $d\neq
1$) of the Virasoro.

Graded TCFT models occur as the topological sigma-models with
a Calabi -- Yau manifold of (complex) dimension $d$ as the
target space [2, 46, 57]. They are nonresonant for even $d$.
Particularly, for $d=2$ one obtains the K3-models where the
primary correlators are constant. For $d>2$ these are not
constant because of instanton corrections [46, 47, 57]. As it was
explained in [57], finding of these primary correlators for the
Calabi -- Yau models (and, therefore, graded solutions of
WDVV) could be a crucial point in solving the problem of
mirror symmetry.

The compatible pair of the Poisson brackets (3.2) and (3.37)
generates an integrable hierarchy of PDE for a non-resonant
graded TCFT using the standard machinery of the bihamiltonian
formalism [52]
$$\partial_{T^{\alpha ,p}}t^\beta =\{ t^\beta (X),\hat H_{\alpha ,p}\} =
\{ t^\beta (X), \hat H_{\alpha ,p-1}\}_1^{\hat{}}.\eqno(3.39)$$
Here the Hamiltonians have the form
$$\hat H_{\alpha ,p} = \int \hat h_{\alpha ,p+1}dX,\eqno(3.40a)$$
$$\hat h_{\alpha ,p+1 }=[{d+1\over 2} -q_\alpha + p]^{-1}
 h_{\alpha ,p+1}(t)
+\epsilon^2 \Delta\hat h_{\alpha ,p+1}(t, \partial_Xt,
\dots , \partial_X^pt;\epsilon^2)\eqno(3.40b)$$
where $\Delta\hat h_{\alpha ,p+1}$ are some polynomials determined
by (3.39). They are graded-homogeneous of degree 2 where deg$\partial^k_Xt
=k$, deg$\epsilon = -1$. The small dispersion parameter $\epsilon$ also
plays the role of the string coupling constant.
It is clear that the hierarchy (3.1) is the
zero-dispersion limit of this hierarchy. For $n=1$ using the
pair (3.2) and (3.38) one immediately obtains the KdV hierarchy.
Note that this describes the topological gravity. It would be
interesting to investigate relation of the hierarchies
determined by the pair (3.2) and (3.37) to a nonperturbative (i.e.,
for all genera) description of the Calabi -- Yau models
(especially, of the K3 models) coupled to gravity.
For a model with constant correlators (for a graded
Fr\"obenius algebra $A$) the first nontrivial equations of the
hierarchy are
$$\partial_{T^{\alpha ,1}}{\bf t} = e_\alpha {\bf t}{\bf t}_X +
{2\epsilon^2\over 3-d}e_\alpha e_n{\bf t}_{XXX}. \eqno(3.41)$$

For non-graded TCFT models it could be of interest to find nonlinear
analogues of the cocycle (3.37). These should be differential geometric
Poisson brackets of the third order [58, 18] of the form
$$\{ t^\alpha (X),t^\beta (Y)\}_1^{\hat{}}
= \{ t^\alpha (X),t^\beta (Y)\}_1 +$$
$$\epsilon^2\{ g^{\alpha\beta}(t(X))\delta '''(X-Y) +
b^{\alpha\beta}_\gamma (t(X))t^\gamma_X\delta ''(X-Y)+$$
$$[f^{\alpha\beta}_\gamma (t(X))t_{XX}^\gamma +h_{\gamma\delta}^{\alpha\beta}
(t(X))t_X^\gamma t_X^\delta ]\delta '(X-Y)+$$
$$[p^{\alpha\beta}_\gamma (t)t_{XXX}^\gamma +
q_{\gamma\delta}^{\alpha\beta}(t)t_{XX}^\gamma t_X^\delta
+r_{\gamma\delta\lambda}^{\alpha\beta}(t)t_X^\gamma t_X^\delta
t_X^\lambda ]\delta (X-Y)\}.\eqno(3.42)$$
I recall (see [58, 18]) that the form (3.42) of the Poisson bracket should be
invariant with respect to nonlinear changes of coordinates in the
manifold $M$. This implies that the leading term $g^{\alpha\beta}(t)$
transforms like a metric (may be, degenerate) on the cotangent bundle $T_*M$,
$b^{\alpha\beta}_\gamma (t)$ are contravariant components of a connection on
$M$ etc. The Poisson bracket (3.42) is assumed to be compatible with (3.2).
Then the compatible pair (3.2), (3.42) of the Poisson brackets generates an
integrable hierarchy of the same structure (3.39), (3.40).
\medskip
{\bf 4. Examples}.
\medskip
I start with the most elementary examples of solutions of WDVV for
$n=2$. Only massive solutions are of interest here (a 2-dimensional
nilpotent Fr\"obenius algebra has no nontrivial deformations). The
Darboux -- Egoroff equations in this case are linear. I consider only
TCFT case (the similarity reduction of WDVV). Let us redenote the
coupling constants
$$t^1 = u,~t^2 = \rho .\eqno(4.1)$$
For $d\neq 1$ the
primary free energy $F$ has the form
$$F = {1\over 2}\rho u^2 + {g\over a(a+2)}\rho^{a+2},\eqno(4.2)$$
$$a={1+d\over 1-d} \eqno(4.3)$$
$g$ is an arbitrary constant. The second term in the formula for the
free energy should be understood as
$${g\over a(a+2)}\rho^{a+2}= \int\int\int g(a+1)\rho^{a-1}.$$
The linear system (3.5) can be solved via Bessel functions [39]. Let me
give an example of equations of the hierarchy (3.1) (the $T=T^{1,1}$-flow)
$$u_T+uu_X+g\rho^a\rho_X  =0 \eqno (4.4a)$$
$$\rho_T+(\rho u)_X = 0.\eqno(4.4b)$$
These are the equations of isentropic motion of one-dimensional fluid with
the dependence of the pressure on the density of the form
$p={g\over a+2}\rho^{a+2}$. The Poisson structure (3.2) for these equations
was proposed in [37].
For $a=0$ (equivalently $d=-1$) the system coincides with the
equations of  waves on shallow water (the dispersionless limit [59] of the
nonlinear Schr\"odinger equation (NLS)).

For $d=1$ the primary free energy has the form
$$F = {1\over 2}\rho u^2 + ge^\rho .\eqno(4.5)$$
This coincides with the free energy of the topological sigma-model
with  $CP^1$ as the target space. Note that this can be obtained
from the same solution of the Darboux -- Egoroff system as the
semiclassical limit of the NLS (the case $d=-1$ above) for different
choices of the eigenfunction $\psi_{1i}$ (in the notations of (1.24)).
The corresponding $T=T^{2,0}$-system of the hierarchy (3.1) reads
$$u_T=g(e^\rho )_X$$
$$\rho_T=u_X.$$
Eliminating $u$ one obtains the long wave limit
$$\rho_{TT} =g(e^\rho )_{XX} \eqno(4.6)$$
of the Toda system
$${\rho_n}_{tt} = e^{\rho_{n+1}} -2e^{\rho_n} + e^{\rho_{n-1}}. \eqno(4.7)$$
(The 2-dimensional version of (4.6)  was obtained in the formalism of
Whitham-type equations in
[44].) It would be interesting to prove that the nonperturbative free
energy of the $CP^1$-model coincides with the $\tau$-function of the
Toda hierarchy.

Example 2. Topological minimal models. I consider here the $A_n$-series
models only. The Fr\"obenius manifold $M$ here is the set of all
polinomials ({\it Landau -- Ginsburg superpotentials}) of the form
$$M = \{ w(p) = p^{n+1} +a_1p^{n-1}+\dots +a_n|~ a_1,\dots ,a_n\in {\bf
C}\} .\eqno(4.8)$$
For any $w\in M$ the Fr\"obenius algebra $A=A_w$ is the algebra of
truncated polynomials
$$A_w ={\bf C}[p]/(w'(p)=0) \eqno(4.9)$$
(the prime means derivative with respect to $p$) with the invariant
inner product
$$<f,g> = {\rm res}_{p=\infty}{f(p)g(p)\over w'(p)}. \eqno(4.10)$$
The algebra $A_w$ is semisimple if the polynomial $w'(p)$ has simple
roots. The canonical coordinates (1.15) $u^1,\dots ,u^n$ are the critical
values of the polynomial $w(p)$
$$u^i = w(p_i),~{\rm where}~ w'(p_i) = 0, ~i=1,\dots ,n. \eqno(4.11)$$
Let us take the following diagonal metric on $M$
$$\sum_{i=1}^n \eta_{ii}(u)(du^i)^2,~~ \eta_{ii}(u) = [w''(p_i)]^{-1}.
\eqno(4.12)$$
It can be proved that this is a flat Egoroff metric on $M$.
The correspondent flat coordinates on $M$ have the form
$$t^\alpha = -{n+1\over n-\alpha +1}{\rm res}_{p=\infty}w^{{n-\alpha +1
\over n+1}}(p)dp, ~\alpha =1 ,\dots ,n. \eqno(4.13)$$
The metric (4.12) in these coordinates has the constant form
$$\sum_{i=1}^n \eta_{ii}(u)(du^i)^2 = \eta_{\alpha\beta} dt^\alpha
dt^\beta ,~~ \eta_{\alpha\beta} = \delta_{n+1,\alpha + \beta}.
\eqno(4.14)$$
The ortonormal basis in $A_w$ with respect to this metric consists of
the polynomials $\phi_1(p)$,... , $\phi_n(p)$ of degrees 0, 1, ... ,
$n-1$ resp. where
$$\phi_\alpha (p) = {d\over dp}[w^{{\alpha\over n+1}}]_+, ~~\alpha =
1,\dots ,n, \eqno(4.15)$$
Here $[~~]_+$ means the polynomial part of the power series in $p$.
This is a TCFT model with the charges and dimension
$$q_\alpha = {\alpha -1\over n+1}, ~d=q_n={n-1\over n+1}.\eqno(4.16)$$

In fact one obtains a $n$-parameter family of TFT models with the same
canonical coordinates $u^i$ of the form (4.11) where
$$\eta_{ii}(u) \mapsto \eta_{ii}(u,c) = [w''(p_i)]^{-1}[\sum c_\alpha
\phi_\alpha (p_i)]^2, \eqno(4.17a)$$
$$t^\alpha \mapsto t^\alpha (c) = -{n+1\over n-\alpha
+1}{\rm res}_{p=\infty}w^{{n-\alpha +1\over n+1}}(p) [\sum c_\gamma \phi_\gamma
(p)]dp \eqno(4.17b)$$
depending on arbitrary parameters $c_1$, ..., $c_n$. This reflects the
ambiguity in the choice of the solution $\psi_{i1}$ in the formulae
(1.24). These models are conformal invariant if only one of the coefficients
$c_\gamma$
is nonzero.

The corresponding hierarchy of the systems of hydrodynamic type (3.1)
coincides with the dispersionless limit of the Gelfand -- Dickey
hierarchy for the scalar Lax operator of order $n+1$. This essentialy
follows
from [8, 11]. I recall that the Gelfand -- Dickey hierarchy for an
operator
$$L= \partial^{n+1} + a_1(x)\partial ^{n-1} +\dots + a_n(x)$$
$$\partial = d/dx$$
has the form
$$\partial_{t^{\alpha ,p}}L = c_{\alpha ,p}[L,[L^{{\alpha\over n+1}+p}]_+],
{}~\alpha =1,\dots ,n,~p=0,1,\dots  \eqno(4.18)$$
for some constants $c_{\alpha ,p}$. Here $[~]_+$ denotes differential
part of the pseudodifferential operator. The dispersionless limit of
the hierarchy is defined as follows: one should substitute
$$x\mapsto \epsilon x=X, ~t^{\alpha ,p}\mapsto \epsilon t^{\alpha ,p} =
T^{\alpha ,p}\eqno(4.19)$$
and tend $\epsilon$ to zero. The dispersionless limit of
$\tau$-function of the hierarchy is defined [11, 53-54, 61] as
$$\log\tau_{{\rm dispersionless}}(T) = \lim_{\epsilon\to 0} \epsilon^{-2}
\log\tau (\epsilon t).\eqno(4.20)$$

Modified minimal model (4.17) is related to the same Gelfand --
Dickey hierarchy with the following modification of the $L$-operator
$$L\mapsto \tilde L=\sum c_\gamma [L^{{\gamma\over n+1}}]_+.\eqno(4.21)$$

The linear equation (3.5) for the minimal model can be solved in the form
[39]
$$h_\alpha (t;\lambda ) =-{n+1\over \alpha}{\rm res}_{p=\infty}
w^{{\alpha\over n+1}}{}_1F_1(1;1+{\alpha\over n+1};\lambda
w(p))dp.\eqno(4.22)$$
Here ${}_1F_1(a;c;z)$ is the Kummer (or confluent hypergeometric)
function [35]
$${}_1F_1(a;c;z)=\sum_{m=o}^\infty {(a)_m\over (c)_m}{z^m\over m!},
\eqno(4.23a)$$
$$(a)_m = a(a+1)\dots (a+m-1).\eqno(4.23b)$$
The generating function (3.11) has the form
$$V_{\alpha\beta}(t;\lambda ,\mu ) =(\lambda + \mu )^{-1}
[\eta^{\mu\nu}({\rm res}_{p=\infty}w^{{\alpha\over n+1}-1}
{}_1F_1(1;{\alpha\over n+1};\lambda w(p))\phi_\mu (p)dp)\times$$
$$({\rm res}_{p=\infty}w^{{\beta\over n+1}-1}{}_1F_1(1;
{\beta\over n+1};\mu w(p))\phi_\nu (p)dp)-\eta_{\alpha\beta}].
\eqno(4.24)$$
{}From this one obtains formulae for the $\tau$-function.

Example 3. $M_{g;n_0,\dots ,n_m}$-models [13, 14].
Let $M=M_{g;n_0,\dots ,n_m}$  be a moduli space of dimension
$$n=2g+n_0+\dots +n_m+2m\eqno(4.25)$$
of sets
$$(C;\infty_0,\dots ,\infty_m; w; k_0,\dots ,k_m;
a_1,\dots ,a_g,b_1,\dots ,b_g)\in M_{g;n_0,\dots ,n_m}\eqno(4.26)$$
where $C$ is a Riemann surface with marked points $\infty_0$, ...,
$\infty_m$, and a marked meromorphic function
$$w:C\to CP^1,~~w^{-1}(\infty ) = \infty_0\cup\dots ,
\cup\infty_m \eqno(4.27)$$
having a degree $n_i+1$ near the point $\infty_i$, and a marked symplectic
basis
$a_1,\dots ,a_g$, $b_1,\dots ,b_g\in H_1(C,{\bf Z})$, and marked
branches of roots
of $w$ near $\infty_0$, ..., $\infty_m$ of the orders $n_0+1$, ..., $n_m+1$
resp.,
$$k_i^{n_i+1}(P) = w(P), ~P~{\rm near}~\infty_i.\eqno(4.28)$$
(This is a connect manifold as it follows from [56].) We need the critical
values of $w$
$$u^j=w(P_j),~dw|_{P_j}=0,~j=1,\dots ,n\eqno(4.29)$$
(i.e. the ramification points of the Riemann surface (4.27)) to be local
coordinates in  open domains in $M$ where
$$u^i\neq u^j~{\rm for}~i\neq j\eqno(4.30)$$
(due to the Riemann existence theorem). Another assumption is that the
one-dimensional affine group acts on $M$ as
$$(C;\infty_0,\dots ,\infty_m;w;\dots )\mapsto
(C;\infty_0,\dots ,\infty_m;aw+b;\dots )\eqno(4.31a)$$
$$u^i\mapsto au^i+b,~ i=1,\dots ,n.\eqno(4.31b)$$
Let $dp$ be the normalized Abelian differential of the second kind on $C$
with a double pole at $\infty_0$
$$dp = dk_0 +~{\rm regular~terms}\eqno(4.32a)$$
$$\oint_{a_i} dp=0,~i=1,\dots ,g.\eqno(4.32b)$$
Using $u^i$ as the canonical coordinates (1.15) I define a flat
Egoroff metric on
$M$ by the formula
$$ds^2=\sum_{i=1}^n\eta_{ii}(u)(du^i)^2,\eqno(4.33a)$$
$$\eta_{ii}(u) = {\rm res}_{P_i}{(dp)^2\over dw}.\eqno(4.33b)$$
It can be extended globaly on $M$. The corresponding flat coordinates are
$$t^{i;\alpha}=-{n_i+1\over n_i-\alpha +1}{\rm res}_{\infty_i}
k_i^{n_i-\alpha +1}dp,~
i=0,\dots ,m,~\alpha =1,\dots ,n_i;\eqno(4.34a)$$
$$p^i= {\rm v.p.}\int_{\infty_0}^{\infty_i}dp =
\lim_{Q\to\infty_0}
(\int_Q^{\infty_i}dp + k_0(Q)),~i=1,\dots ,m;\eqno(4.34b)$$
$$q^i = -{\rm res}_{\infty_i} wdp,~i=1,\dots ,m;\eqno(4.34c)$$
$$r^i = \oint_{b_i}dp,~s^i=-{1\over 2\pi i}\oint_{a_i}wdp,~
i=1,\dots ,g.\eqno(4.34d)$$
The metric (4.33) in the coordinates has the following form
$$\eta_{t^{i;\alpha}t^{i;\beta}} = {1\over n_i+1}\delta_{ij}
\delta_{\alpha +\beta ,n_i+1}\eqno(4.35a)$$
$$\eta_{p^iq^j} = \delta_{ij}\eqno(4.35b)$$
$$\eta_{r^is^j} = \delta_{ij},\eqno(4.35c)$$
other components of the $\eta$ vanish. The unity vector field is a
unit vector along the coordinate $t^{0;1}$.

{\bf Proposition 4.1.} {\it The flat metric (4.35) is well-defined globaly on
$M$ and the flat coordinates (4.34) are globaly independent analytic functions
on $M$.}

As a consequence we obtain
that the moduli space $M$ is an unramified covering over a domain in
${\bf C}^n$ (see [13, 14]).

Let us introduce primary differentials on $C$ (or on a universal covering
$\tilde C$ of $C\setminus\infty_0\cup\dots\cup\infty_m$) of the form
$$\phi_{t^A}=\partial_{t^A}(pdw)_{w={\rm const}}\eqno(4.36)$$
where
$$p(P)=\int_{Q_0}^Pdp,\eqno(4.37a)$$
$$Q_0\in C,~w(Q_0)=0,\eqno(4.37b)$$
$t^A$ is one of the flat coordinates (4.34). Note that the definition (4.36) of
the primary differentials can be rewritten as
$$\phi_{t^A} = -\partial_{t^A}(wdp)_{p={\rm const}}\eqno(4.38)$$
where the multivalued coordinate $p$ on $C$ is defined in (4.37). So $w(p)$
plays the role of the Landau -- Ginsburg superpotential for the
$M_{g;n_0,\dots ,n_m}$-models. More explicitly, $\phi_{t^{i;\alpha}}$
is a normalized Abelian differential of the second kind with a pole in
$\infty_i$,
$$\phi_{t^{i;\alpha}} = -{1\over\alpha}dk_i^\alpha
+~{\rm regular~terms~~~near~}\infty_i,$$
$$\oint_{a_j}\phi_{t^{i;\alpha}}=0;\eqno(4.39a)$$
$\phi_{p^i}$ is a normalized Abelian differential of the second kind on $C$
with
a pole only at $\infty_i$ with the principal part of the form
$$\phi_{p^i} = dw +~{\rm regular~terms~~~~near~}\infty_i,$$
$$\oint_{a_j}\phi_{p^i} = 0;\eqno(4.39b)$$
$\phi_{q^i}$ is a normalized Abelian differential of the third kind with
simple poles at $\infty_0$ and $\infty_i$ with residues -1 and +1 resp.;

\noindent $\phi_{r^i}$ is a normalized multivalued
differential on $C$ with increments
along the cycles $b_i$ of the form
$$\phi_{r^i} (P+b_j)-\phi_{r^i}(P)
= -\delta_{ij}dw,$$
$$\oint_{a_j}\phi_{r^i} = 0;\eqno(4.39c)$$
$\phi_{s^i}$ are the basic holomorphic differentials\footnote{$^*$}{1-form
$pdw$ was used by Novikov and Veselov in their theory of algebro-geometric
Poisson brackets [60]. The coordinates $s^i$ are the algebro-geometric
action variables of [60]. In [60]
it also was an important point that derivatives
$\partial_{s^i}(pdw)$ are the normalized holomorphic differentials.}
on $C$ normalized by the condition
$$\oint_{a_j}\phi_{s^i} = 2\pi i\delta_{ij}.\eqno(4.39d)$$
The inner product (4.33) in terms of the primary differentials $\phi_{t^A}$
reads
$$\eta_{AB} = \sum_{i=1}^n{\rm res}_{P_i}{\phi_{t^A}\phi_{t^B}\over dw}.
\eqno(4.40)$$
The structure functions $c_{ABC}(t)$ can be calculated as
$$c_{ABC}(t) =\sum_{i=1}^n{\rm res}_{P_i}
{\phi_{t^A}\phi_{t^B}\phi_{t^C}\over dwdp}.\eqno(4.41)$$
Extension of the Fr\"obenius structure on all the moduli space $M$ is
given by the condition that the differential
$${\phi_{t^A}\phi_{t^B}-c_{AB}^C\phi_{t^C}dp\over dw}\eqno(4.42)$$
is holomorphic for $|w|<\infty$. The Fr\"obenius algebra on $T_tM$ will be
nilpotent for Riemann surfaces $w:C\to CP^1$ with more than double branch
points. This is a conformal invariant Fr\"obenius manifold with the
dimension
$$d={n_0-1\over n_0+1}\eqno(4.43a)$$
and charges
$$q_{t^{i;\alpha}}={\alpha\over n_i+1} - {1\over n_0+1}\eqno(4.43b)$$
$$q_{r^i}=q_{p^i}={n_0\over n_0+1}\eqno(4.43c)$$
$$q_{s^i} = q_{q^i} = -{1\over n_0+1}.\eqno(4.43d)$$
For the particular case $g=m=0$ we obtain the Fr\"obenius manifolds
of the minimal models (the previous example). For $g=0$, $m>0$ we obtain
models with rational functions as superpotentials.

The generating functions $h_{t^A}(t;\lambda )$ (3.4) have the form
$$h_{t^{i;\alpha}} (t;\lambda ) =-{n_i+1\over \alpha}{\rm res}_{p=\infty_i}
k_i^\alpha {}_1F_1(1;1+{\alpha\over n_i+1};\lambda w(p))dp.\eqno(4.44a)$$
$$h_{p^i}={\rm v.p.}\int_{\infty_0}^{\infty_i}e^{\lambda w}dp\eqno(4.44b)$$
$$h_{q^i} = {\rm res}_{\infty_i}{e^{\lambda w}-1\over\lambda}dp
\eqno(4.44c)$$
$$h_{r^i} = \oint_{b_i}e^{\lambda w}dp\eqno(4.44d)$$
$$h_{s^i} ={1\over 2\pi i}\oint_{a_i}pe^{\lambda w}dw.\eqno(4.44e)$$

{\bf Remark.} Integrals of the form (4.44) seem to be interesting functions
on the moduli space of the form $M_{g;n_0,\dots ,n_m}$. A simplest example of
such an integral for a family of elliptic curves reads
$$\int_0^\omega e^{\lambda\wp (z)}dz\eqno(4.45)$$
where $\wp (z)$ is the Weierstrass function with periods $2\omega$, $2\omega
'$.
For real negative $\lambda$ a degeneration of the elliptic curve ($\omega
\to\infty$) reduces (4.45) to the standard probability integral
$\int_0^\infty e^{\lambda x^2}dx$. So the integral (4.45) is an analogue
of the probability integral as a function on $\lambda$ and on moduli of
the elliptic curve. I recall that dependence on these parameters is specified
by the equations (3.5), (2.17).

Gradients of this functions on the moduli space $M$ have the form
$$\partial_{t^A}h_{t^{i;\alpha}} = {\rm res}_{\infty_i}
k_i^{\alpha - n_i-1}
{}_1F_1(1;{\alpha\over n_i+1};\lambda w(p))\phi_{t^A},\eqno(4.46a)$$
$$\partial_{t^A}h_{p^i}=\eta_{t^Ap^i}-\lambda {\rm v.p.}
\int_{\infty_0}^{\infty_i}e^{\lambda w}\phi_{t^A}\eqno(4.46b)$$
$$\partial_{t^A}h_{q^i} =
{\rm res}_{\infty_i} e^{\lambda w}\phi_{t^A}\eqno(4.46c)$$
$$\partial_{t^A}h_{r^i}=\eta_{t^Ar^i}-\lambda\oint_{b_i}e^{\lambda w}
\phi_{t^A}\eqno(4.46d)$$
$$\partial_{t^A}h_{s^i}={1\over 2\pi i}\oint_{a_i}
e^{\lambda w}\phi_{t^A}.\eqno(4.46e)$$
The generating function $V_{\alpha\beta}(t;\lambda ,\mu )$ of coefficients
of the $\tau$-function (3.13) can be calculated via inner products (w.r.t. the
matrix (4.35)) of (4.46). Particularly, for a part of the Hessian of the
primary
free energy $F(t)$ (a function on $M$) one obtains [13, 14]
$${\partial^2F\over\partial s^i\partial s^j} = -\tau_{ij}
= -\oint_{b_j}\phi_{s^i}.\eqno(4.47)$$
This is nothing but the matrix of periods of holomorphic differentials on
$M$. Other second derivatives of $F$ also turn out to be certain periods of
some Abelian differentials on $C$.

{\bf Conclusion.} {\it WDVV is a universal system of integrable
differential equations
for periods of Abelian differentials on Riemann surfaces.}

I recall that this system is a high-order analogue of the
Painlev\'e-VI equation (i.e. equations of isomonodromy deformations of (2.15)).
To specify the solution of WDVV one needs to find the monodromy matrix of
the linear operator (2.15) for the eigenfunctions of the form (4.44). I will
do it in a forthcoming publication.

We obtain the following picture of \lq\lq Painlev\'e uniformisation" of
the moduli spaces $M_{g;n_0,\dots ,n_m}$: (1) a global system of analytic
coordinates on $M_{g;n_0,\dots ,n_m}$; (2) periods of Abelian differentials
on curves $C\in M_{g;n_0,\dots ,n_m}$ are certain high-order Painlev\'e
transcendents as functions of these coordinates.

{\bf Remark.} For any Hamiltonian $H_{A ,p}$ of the form (3.3), (4.44)  one can
construct a differential $\Omega_{A,p}$ on $C$ or on the covering $\tilde C$
with singulariries only at the marked infinite points such that
$${\partial\over\partial u^i}h_{t^A,p}={\rm res}_{P_i}
{\Omega_{A,p}dp\over dw},~ i=1,\dots ,n.\eqno(4.48)$$
See[13] for an explicit form of these differentials (for $m=0$ also see
[14]).
Using these differentials the hierarchy (3.1) can be written in the
Flaschka -- Forest --McLaughlin form [16]
$$\partial_{T^{A,p}}dp = \partial_X\Omega_{A,p}\eqno(4.49)$$
(derivatives of the differentials are to be calculated with $w=$const.).

The matrix $V_{(A,p),(B,q)(t)}$ determines a pairing of these differentials
with values in functions on the moduli space
$$(\Omega_{A,p},\Omega_{B,q})=V_{(A,p),(B,q)}(t)\eqno(4.50)$$
Particularly, the primary free energy $F$ as a function on $M$ can be written
in the form [13, 14]
$$F=-{1\over 2}(pdw,pdw).\eqno(4.51)$$
Note that the differential $pdw$ can be written in the form
$$pdw =\sum{n_i+1\over n_i+2}\Omega_{\infty_i}^{(n_i+2)}+
\sum t^A\phi_{t^A}\eqno(4.52)$$
where
$\Omega_{\infty_i}^{(n_i+2)}$ is the Abelian differential of the second kind
with a pole at $\infty_i$ of the form
$$\Omega_{\infty_i}^{(n_i+2)} = dk_i^{n_i+2} +
{}~{\rm regular~terms~~~~near~}\infty_i.\eqno(4.53)$$
For the pairing (4.50) one can obtain from [44] the following formula
$$(f_1dw,f_2dw)={1\over 2}
\int\int_C(\bar\partial f_1\partial f_2
+\partial f_1\bar\partial f_2)\eqno(4.54)$$
where the differentials $\partial$ and $\bar\partial$ along the Riemann surface
should be understood in the distribution sense. The meromorphic differentials
$f_1dw$ and $f_2dw$ on the covering $\tilde C$ should be considered as
piecewise
meromorphic differentials on $C$ with jumps on some cuts.

The corresponding hierarchy (3.1) is obtained by averaging along invariant tori
of a family of $g$-gap solutions of a KdV-type hierarchy related to a matrix
operator $L$ of the matrix order $m+1$ and of orders $n_0$, ..., $n_m$
in $\partial /\partial x$.
The example $m=0$ (the averaged Gelfand -- Dickey hierarchy)
was considered in more details in [14]. The Poisson bracket (3.2) is a result
of semiclassical limit (or averaging) [18-21] of the first haniltonian
structure of the Gelfand -- Dickey hierarchy; averaging of the second
hamiltonian structure (the classical $W$-algebra) gives the Poisson
structure (3.21). Therefore, (3.21) can be considered as semiclassical
limit of classical $W$-algebras. The corresponding flat metric (3.22)
on the moduli space $M_{g;n_0,\dots ,n_m}$
is well-defined on a subset of Riemann
surfaces having $w=0$ a non-ramifying point.

Also for $g+m>0$ one needs to extend the
KdV-type hierarchy to obtain (3.1) (see [13-14]). To explain the nature of such
an extension let us consider the simplest example of $m=0$, $n_0=1$. The moduli
space $M$ consists of hyperelliptic curves of genus $g$ with marked homology
basis
$$y^2 = \prod_{i=1}^{2g+1}(w-w_i).\eqno(4.55)$$
This parametrizes the family of $g$-gap solutions of the KdV. The $L$ operator
has the well-known form
$$L= -\partial^2_x+u.\eqno(4.56)$$
In real smooth periodic case $u(x+T)=u(x)$ the quasimomentum $p(w)$ is
defined by the formula
$$\psi (x+T,w) = e^{ip(w)T}\psi (x,w)\eqno(4.57)$$
for a solution $\psi (x,w)$ of the equation
$$L\psi = w\psi \eqno(4.58)$$
(the Bloch -- Floquet eigenfunction). The differential $dp$ can be extended
onto the family of all (i.e. quasiperiodic complex meromorphic)
$g$-gap operators (4.55) as a
normalized Abelian differential of the second kind with a double pole at the
infinity $w=\infty$. (So the above superpotential (4.38) has the sense of
the Bloch dispersion law, i.e. the dependence of the energy $w$ on
the quasimomentum $p$.) The Hamiltonians of the KdV hierarchy can be
obtained as coefficients of expansion of $dp$ near the infinity. To
obtain a complete family of conservation laws of the averaged hierarchy
(3.1) one needs to extend the family of the KdV integrals by adding
nonlocal functionals of $u$ of the form
$$\oint_{a_i}w^kdp,~~\oint_{b_i}w^{k-1}dp,~k=1,2,\dots .\eqno(4.59)$$

As in (4.17) one can deform the above Fr\"obenius structure on the moduli
space $M=M_{g;n_0,\dots ,n_m}$ by changing the differential $dp$,
$$dp\mapsto \tilde{dp}=\sum c_A\phi_{t^A}\eqno(4.60)$$
for arbitrary constant coefficients. (The deformed Fr\"obenius structure
genericaly is well-defined only on a subset of $M$.) Particularly, if
$\tilde{dp}$ is a differential of the third kind on $C$ then the
\lq\lq dimension" $d$ of this model always equals 1. The corresponding
hierarchy (3.1) is obtained by averaging a Toda-type system.

Here I consider the simplest example of such a deformation. Let us
consider the 3-dimensional family $M$ of elliptic curves
$$y^2 = 4(w-c)^2 - g_2(w-c) -g_3 =
4(w-c-e_1)(w-c-e_2)(w-c-e_3)\eqno(4.61)$$
with ordered roots $e_1$, $e_2$, $e_3$. It is convenient to use the
Weierstrass uniformization of (4.63)
$$w=\wp (z)+c\eqno(4.62a)$$
$$y=\wp '(z)\eqno(4.62b)$$
(I will use the standard notations [35] of the theory of elliptic
functions). Let us use the holomorphic differential
$$dp = {\pi i dz\over\omega}\eqno(4.63)$$
to construct a Fr\"obenius structure on $M$ (here $\wp (\omega )
= e_1$). The corresponding Landau -- Ginsburg superpotential is the
Weierstrass function (4.62a) where one should substitute $z=\omega
p/\pi i$. The flat coordinates $t^1$, $t^2$, $t^3$ for the superpotential
read
$$t^1 = -c +{\eta\over\omega}\eqno(4.64a)$$
$$t^2 = -1/\omega\eqno(4.64b)$$
$$t^3 = 2\pi i\tau ~~{\rm where~} \tau = \omega '/\omega ,\eqno(4.64c)$$
$\wp (\omega ')=e_3$, $\eta = -\int_0^\omega\wp (z)dz$. The charges of this
manifold are $q_0=0$, $q_1={1\over 2}$, $q_2 = d = 1$.

{\bf Remark.} The above models with $m=0$, $g>0$ can be obtained [39]
in a semiclassical description of correlators of multimatrix models
(at the tree-level approximation for small couplings they correspond to
various self-similar solutions of the hierarchy (3.1)) as functions of the
couplings after passing through a point of gradient catastrophe. The idea of
such a description is originated in the theory of a dispersive analogue
of shock waves [32]; see also [18].

More general algebraic-geometrical examples of solutions of WDVV were
constructed in [44]. In these examples $M$ is a moduli space of
Riemann surfaces of genus $g$ with a marked normalized Abelian differential
of the second kind $dw$ with poles at marked points and with fixed
$b$-periods
$$\oint_{b_i}=B_i, ~i=1,\dots ,g.$$
For $B_i=0$ one obtains the above Fr\"obenius structures on
$M_{g;n_0,\dots ,n_m}$. Unfortunately, for $B\neq 0$ the Fr\"obenius
structures of [44] does not admit a conformal invariance.

\medskip
{\bf Acknowledgments.} I wish to thank E.Witten and C.Vafa for
instructive and stimulating discussions. I am acknowledged
to INFN, Sez. di Napoli, where this paper was completed, for support.

\vfill\eject

{\bf References}
\medskip

\item{1.} F. Br\'ezin and V. Kazakov, {\sl Phys. Lett.} {\bf B 236} (1990)
144.\item{}
M. Douglas and S. Shenker, {\sl Nucl. Phys.} {\bf B 335} (1990) 635. \item{}
D.J. Gross and A. Migdal, {\sl Phys. Rev. Lett.} {\bf 64} (1990) 127.\item{}
D.J. Gross and A. Migdal, {\sl Nucl. Phys.} {\bf B 340} (1990) 333. \item{}
T. Banks, M. Douglas, N. Seiberg, and S. Shenker, {\sl Phys. Lett.} {\bf B
238} (1990) 279. \item{}
M. Douglas, {\sl Phys. Lett.} {\bf B 238} (1990) 176.
\medskip
\item{2.} E. Witten, {\sl Commun. Math. Phys.} {\bf 117} (1988) 353; {\bf 118}
(1988) 411.
\medskip
\item{3.} E. Witten, {\sl Nucl. Phys.} {\bf B 340} (1990) 281.
\medskip
\item{4.} R. Dijkgraaf and E. Witten, {\sl Nucl. Phys.} {\bf B 342} (1990)
486.
\medskip
\item{5.} J. Distler, {\sl Nucl. Phys.} {\bf B 342} (1990) 523.
\medskip
\item{6.} K. Li, {\it Topological gravity and minimal matter},
CALT-68-1662, August
1990; {\it Recursion relations in topological gravity with minimal matter},
CALT-68-1670, September 1990.
\medskip
\item{7.} E. Martinec, {\sl Phys. Lett.} {\bf B 217} (1989) 431; {\it
Criticality,
Catastrophe and Compactifications}, V.G. Knizhnik memorial volume, 1989.\item{}
C. Vafa and N.P. Warner, {\sl Nucl. Phys.} {\bf B 324} (1989) 427. \item{}
S. Cecotti, L. Girardello and A. Pasquinucci, {\sl Nucl. Phys.} {\bf B
328} (1989) 701;\item{}
S. Cecotti, L. Girardello and A. Pasquinucci, {\sl Int. J. Mod. Phys.}
{\bf A 6} (1991) 2427.
\medskip
\item{8.} R. Dijkgraaf. E. Verlinde and H. Verlinde, {\sl Nucl. Phys.} {\bf B
352} (1991) 59;\item{}
{\it Notes on topological string theory and 2D quantum gravity}, PUPT-1217,
IASSNS-HEP-90/80, November 1990.
\medskip
\item{9.} C. Vafa, {\sl Mod. Phys. Lett.} {\bf A 6} (1991) 337.
\medskip
\item{10.} B. Blok and A. Varchenko, {\sl Int. J. Mod. Phys.} {\bf A7}
(1992) 1467.
\medskip
\item{11.} I. Krichever, {\sl Comm. Math. Phys.} {\bf 143} (1992) 415.
\medskip
\item{12.} I. Krichever, {\it Whitham theory for integrable systems and
topological field theories}. To appear in {\sl Proceedings of Summer Cargese
School}, July 1991.
\medskip
\item{13.} B. Dubrovin, {\it Differential geometry of moduli spaces and its
application to soliton equations and to topological conformal field theory},
Preprint No. 117 of Scuola Normale Superiore, Pisa, November 1991.
\medskip
\item{14.} B. Dubrovin, {\sl
Comm. Math. Phys.} {\bf 145} (1992) 195.
\medskip
\item{15.} G.B. Whitham, {\it Linear and Nonlinear Waves},
Wiley Intersci., New York
- London - Sydney, 1974.\item{}
S. Yu. Dobrokhotov and V.P. Maslov, {\it Multiphase asymptotics of nonlinear
PDE
with a small parameter}, {\sl Sov. Sci. Rev.: Math. Phys. Rev.} {\bf 3}
(1982) 221.
\medskip
\item{16.} H. Flaschka, M.G. Forest and D.W. McLaughlin, {\sl Comm. Pure Appl.
Math.} {\bf 33} (1980) 739.
\medskip
\item{17.} I. Krichever, {\sl Funct. Anal. Appl.} {\bf 22} (1988) 200;
{\sl Russ. Math. Surveys} {\bf 44} (1989) 145.
\medskip
\item{18.} B. Dubrovin and S. Novikov, {\sl Russ. Math. Surveys} {\bf 44}:6
(1989) 35.
\medskip
\item{19.} B. Dubrovin and S. Novikov, {\sl Sov. Math. Doklady} {\bf 27} (1983)
665.
\medskip
\item{20.}
S. Novikov, {\sl Russ. Math. Surveys} {\bf 40}:4 (1985) 85.
\medskip
\item{21.}
B. Dubrovin, {\it Geometry of Hamiltonian  Evolutionary Systems}, Bibliopolis,
Naples 1991.
\medskip
\item{22.} S. Tsarev, {\sl Math. USSR Izvestija} {\bf 36} (1991); {\sl Sov.
Math. Dokl.} {\bf 34} (1985) 534.
\medskip
\item{23.} B. Dubrovin, {\sl Funct. Anal. Appl.} {\bf 24} (1990).
\medskip
\item{24.} B. Dubrovin, {\sl Funct. Anal. Appl.} {\bf 11} (1977) 265.
\medskip
\item{25.} S.P. Novikov (Ed.), {\it The Theory of Solitons: the Inverse Problem
Method}, Nauka, Moscow, 1980. Translation: Plenum Press, N.Y., 1984.
\medskip
\item{26.} V. Zakharov and S. Manakov, {\sl Sov. Phys. JETP} {\bf 42} (1976)
842.
\medskip
\item{27.}
B. Dubrovin, {\sl J. Sov. Math.} {\bf 28} (1985) 20; {\it Theory of operators
and real algebraic geometry}, In: {\sl Lecture Notes in Mathematics} {\bf 1334}
(1988) 42.
\medskip
\item{28.}
B. Dubrovin, I. Krichever and S. Novikov, {\it Integrable Systems}. I.
Encyclopaedia of Mathematical Sciences, vol.4 (1985) 173,
Springer-Verlag.
\medskip
\item{29.}
L.D. Faddeev and L.A. Takhtajan, {\it Hamiltonian Methods in the Theory of
Solitons}. Springer-Verlag, 1987.
\medskip
\item{30.} A.S. Fokas, R.A. Leo, L. Martina, and G. Soliani, {\sl Phys. Lett.}
{\bf A 115} (1986) 329.
\medskip
\item{31.} B.M.McCoy, C.A.Tracy, T.T.Wu, {\sl J. Math. Phys.} {\bf 18} (1977)
1058;
\item{} M.Sato, T.Miwa, and M.Jimbo, {\sl Publ. RIMS} {\bf 14} (1978) 223;
{\bf 15} (1979) 201, 577, 871; {\bf 16} (1980) 531.
\item{} A.Jimbo, T.Miwa, Y.Mori, M.Sato, Physica {\bf 1D} (1980) 80.
\item{} H. Flaschka and A.C. Newell, {\sl Comm. Math. Phys.} {\bf 76} (1980)
65.\item{}
A.R. Its and V. Yu. Novokshenov, {\it The Isomonodromic Deformation
Method in the Theory of Painlev\'e Equations}, {\sl Lecture Notes in
Mathematics}
1191, Springer-Verlag, Berlin 1986.
\medskip
\item{32.} A.V. Gurevich and L.P. Pitaevskii, {\sl Sov. Phys. JETP} {\bf 38}
(1974) 291; {\sl ibid.}, {\bf 93} (1987) 871; {\sl JETP Letters} {\bf 17}
(1973) 193.\item{}
V. Avilov and S. Novikov, {\sl Sov. Phys. Dokl.} {\bf 32} (1987) 366.\item{}
V. Avilov, I. Krichever and S. Novikov, {\sl Sov. Phys. Dokl.} {\bf 32}
(1987) 564.
\medskip
\item{33.} G. Darboux, {\it Le\c{c}ons sur les syst\`emes ortogonaux et les
cordonn\'ees curvilignes}, Paris, 1897. \item{}
D. Th. Egoroff, {\it Collected papers on differential geometry}, Nauka, Moscow
(1970) (in Russian).
\medskip
\item{34.} A. Balinskii and S. Novikov, {\sl Sov. Math. Dokl.} {\bf 32} (1985)
228.
\medskip
\item{35.} W. Magnus, F. Oberhettinger and R.P. Soni, {\it
Formulas and Theorems for
the Special Functions of Mathematical Physics}, Springer-Verlag, Berlin -
Heidelberg - New York, 1966.
\medskip
\item{36.} I.M. Gelfand and L.A. Dickey, {\it
A Family of Hamilton Structures Related
to Integrable Systems}, preprint IPM/136 (1978) (in Russian).\item{}
M. Adler, {\sl Invent. Math.} {\bf 50} (1979) 219.\item{}
I. Gelfand and I. Dorfman, {\sl Funct. Anal. Appl.} {\bf 14} (1980) 223.
\medskip
\item{37.} P.J.Olver, {\sl Math. Proc. Cambridge Philos. Soc.} {\bf 88}
(1980) 71.
\medskip
\item{38.} R.Dijkgraaf, {\it Intersection
Theory, Integrable Hierarchies and Topological Field Theory}, Preprint
IASSNS-HEP-91/91, December 1991.
\medskip
\item{39.} B.Dubrovin, {\sl Nucl. Phys.} {\bf B 379} (1992) 627.
\medskip
\item{40.} S.Cecotti, C.Vafa, {\sl Nucl. Phys.} {\bf B367} (1991) 359.
\medskip
\item{41.} A.R.Its, A.G.Izergin, and V.E.Korepin, {\sl Comm. Math. Phys.}
{\bf 129} (1990) 205.
\medskip
\item{42.} M.Kontsevich, {\sl Funct. Anal. Appl.} {\bf 25} (1991) 50.
\medskip
\item{43.} M.Kontsevich, {\sl Comm. Math. Phys.} {\bf 147} (1992) 1.
\medskip
\item{44.} I.Krichever, {\it The $\tau$-function of the universal Whitham
hierarchy, matrix models and topological field theories}, Preprint
LPTENS-92/18 (hep-th@xxx/9205110).
\medskip
\item{45.}B.Dubrovin, {\it Geometry and integrability of topological -
antitopological fusion}, Preprint INFN-8/92-DSF (hep-th@xxx/9206037),
to appear in {\sl Comm. Math. Phys.}
\medskip
\item{46.} E.Witten, {\sl Surv. Diff. Geom.}{\bf 1} (1991) 243.
\medskip
\item{47.} C.Vafa, {\it Topological Mirrors and Quantum Rings}, Preprint
HUTP-91/A059.
\medskip
\item{48.} E.Witten, {\it Algebraic Geometry Associated with Matrix Models
of Two-Dimensional Gravity}, Preprint IASSNS-HEP-91/74.
\medskip
\item{49.} E.Witten, {\sl Int. J. Mod. Phys.} {\bf A6} (1991) 2775.
\medskip
\item{50.} M.F.Atiyah, {\it Topological Quantum Field Theories}, Publ. Math.
I.H.E.S. {\bf 68} (1988) 175.
\medskip
\item{51.} R.Dijkgraaf, {\it A Geometrical Approach to Two-Dimensional
Conformal
Field Theory}, Ph.D. Thesis (Utrecht, 1989).
\medskip
\item{52.} F.Magri, {\sl J. Math. Phys.} {\bf 19} (1978) 1156.
\medskip
\item{53.} K.Takasaki and T.Takebe, {\it Quasi-classical limit of KP
hierarchy, W-symmetries and free fermions}, Preprint KUCP-0050/92;
{\it W-algebra, twistor, and nonlinear integrable systems}, Preprint
KUCP-0049/92.
\medskip
\item{54.} Y.Kodama, {\sl Phys. Lett.} {\bf 129A} (1988) 223;
{\sl Phys. Lett.} {\bf 147A} (1990) 477; \item{}
Y.Kodama and J.Gibbons, {\sl Phys. Lett.}{\bf 135A} (1989) 167.
\medskip
\item{55.}E.Brezin, C.Itzykson, G.Parisi, and J.-B.Zuber, {\sl Comm.
Math. Phys.} {\bf 59} (1978) 35.
\item{} D.Bessis, C.Itzykson, and J.-B.Zuber, {\sl Adv. Appl. Math.}
{\bf 1} (1980) 109.
\item{} M.L.Mehta, {\sl Comm. Math. Phys.} {\bf 79} (1981) 327;
\item{} S.Chadha, G.Mahoux, and M.L.Mehta, {\sl J.Phys.} {\bf A14}
(1981) 579.
\medskip
\item{56.} S.Natanzon, {\sl Sov. Math. Dokl.} {\bf 30} (1984) 724.
\medskip
\item{57.} E.Witten, {\it Lectures on Mirror Symmetry}, In: {\sl
Proceedings MSRI Conference on Mirror Symmetry}, March 1991, Berkeley.
\medskip
\item{58.} B.Dubrovin and S.Novikov, {\sl Sov. Math. Doklady} {\bf 279} (1984)
294.
\medskip
\item{59.} V.Zakharov, {\sl Funct. Anal. Appl.} {\bf 14} (1980).
\medskip
\item{60.} S.Novikov and A.Veselov, {\sl Sov. Math. Doklady} (1981);
\item{} {\sl Proceedings of Steklov Institute} (1984).
\medskip
\item{61.} P.D.Lax, C.D.Levermore, and S.Venakides, {\it The generation
and propagation of oscillations in dispersive IVPs and their limiting
behaviour}, In: {\sl Important Developments in Soliton Theory 1980 -
1990}, Eds. A.Fokas and V.Zakharov.
\medskip
\hfill September 1992.
\vfill\eject\end